\documentclass[pdflatex,sn-mathphys-num]{sn-jnl}

\usepackage{graphicx}%
\usepackage{multirow}%
\usepackage{amsmath,amssymb,amsfonts}%
\usepackage{amsthm}%
\usepackage{mathrsfs}%
\usepackage[title]{appendix}%
\usepackage{xcolor}%
\usepackage{textcomp}%
\usepackage{manyfoot}%
\usepackage{booktabs}%
\usepackage{algorithm}%
\usepackage{algorithmicx}%
\usepackage{algpseudocode}%
\usepackage{listings}%
\usepackage[inline]{enumitem}
\usepackage{tasks}
\usepackage{float}
\usepackage{siunitx}

\AtBeginDocument{}

\newenvironment{supplement}{%
    \setcounter{table}{0}%
    \renewcommand{\tablename}{}%
    \renewcommand{\thetable}{Supplementary Table \arabic{table}}%
    \setcounter{figure}{0}%
    \renewcommand{\figurename}{}%
    \renewcommand{\thefigure}{Supplementary Figure \arabic{figure}}%
}{}

\begin{document}

\title[Cascading disruptions in natural gas, fertilizers, and crops drive structural food supply vulnerabilities globally]{Cascading disruptions in natural gas, fertilizers, and crops drive structural food supply vulnerabilities globally}


\author*[1]{\fnm{Pavel} \sur{Kiparisov}}\email{kiparisov@iiasa.ac.at}

\author*[1]{\fnm{Christian} \sur{Folberth}}\email{folberth@iiasa.ac.at}


\affil*[1]{\orgdiv{Agriculture, Forestry, and Ecosystem Services (AFE) Research Group, Biodiversity and Natural Resources (BNR) Program}, \orgname{International Institute for Applied Systems Analysis (IIASA)}, \orgaddress{\street{Schlossplatz 1}, \city{Laxenburg}, \postcode{A-2361}, \country{Austria}}}




\abstract{Global food security depends on tightly coupled international supply chains encompassing natural gas, mineral fertilizers, and staple crops. Earlier research has examined the potential consequences of disruptions in each of these domains separately, but not from a systemic perspective. Here we integrate bilateral trade in natural gas, nitrogen, phosphorus, and potassium fertilizers, and eleven staple crops -- accounting for approximately 70\% of plant-based calories -- into a cascading-impact model spanning 208 countries, 20 geopolitical blocs, and the period 1992--2023. Under complete trade isolation, up to 22\% of global caloric consumption would be lost, with a peak in the most recently evaluated years. Structural vulnerabilities vary considerably. Regions largely lacking some segments of the supply chain face near-total crop supply collapse, while few countries can cover the entire nexus through domestic resource endowments and production capacities. Temporal trends highlight a substantial increase in vulnerability globally, most prominently in the EU, with a near two-fold increase since the 1990s. Market power is most concentrated and most volatile in the upstream gas layer and has risen in the fertilizer layers since the 2000s; shocks propagate downstream from these tightening upstream layers, driving the system's fragility. Food stocks provide only limited resilience, with half of humanity living in countries holding stocks lasting fewer than three months. Our results identify upstream supply chains as the structural bottlenecks of the global agrifood system and propose leverage points to enhance resilience.}


%
%
%




\maketitle

\section{Introduction}\label{intro}



Food supply shocks threaten global food security, particularly as supply chains for agronomic inputs, agricultural commodities, and food products have become increasingly interdependent \citep{kummu2020, jia2024}. While their globally integrated trade can support local crop production with vital inputs and buffer local food shortages from hazards, this interconnectedness itself creates structural vulnerabilities. When key trading partners withdraw or critical supply routes close due to conflict, political instability, or infrastructure collapse, import-dependent countries face abrupt supply disruptions with limited alternatives \citep{laber2023, li2023}. Rising geopolitical tensions -- from armed to trade conflicts and the formation of political blocs -- have repeatedly and increasingly fragmented global food trade networks over the past decades \citep{goes2023, gopinath2024, airaudo2025}. In response to extreme events, countries restrict exports to secure domestic supplies, imposing trade barriers that create compounding and cascading disruptions extending far beyond direct rivalry zones \citep{zahoor2023, wef2025}. Accordingly, \citet{quitzow2025} argue that geopolitical tensions, nutrient use efficiency, decarbonization pressures, and food security concerns are deeply intertwined across global fertilizer supply chains, calling for a nexus-based research agenda to address these interdependencies.


Over the past decades, increases in crop production have to a large extent been driven by increasing nutrient inputs, in addition to crop breeding and other agronomic improvements \citep{foley2011} -- a trend that is expected to continue in order to avoid further expansion of cropland \citep{zabel2019}. All three major macronutrients underpinning this intensification -- nitrogen (N), phosphorus (P), and potassium (K) -- are tightly coupled to the fossil fuel and mining sectors and exhibit strong geographic concentration and geopolitical exposure in their trade \citep{brownlie2024, tonelli2024}. Natural gas, for example, is the key source of energy and hydrogen in N fertilizer production, rendering the energy--nitrogen linkage a vital step in fertilizer and, ultimately, food supply \citep{oecd2024}. N fertilizer alone is estimated to currently feed approximately 3.8 billion people, and 1.78 billion people per year are fed from food reliant on either imported fertilizers or imported gas \citep{rosa2023}. The other two macronutrients, P and K, are in turn sourced predominantly from geographically concentrated deposits, with around 60\% of P exports originating from China, Morocco, and Russia and approximately 80\% of K fertilizers from Canada, Russia, Belarus, and China \citep{brownlie2024, oecd2024}.


Recent global shocks have constrained exports from major food-producing countries, disrupted access to fertilizers and energy, and impaired critical trade routes \citep{alexander2023}. Between 2019 and 2024, moderate or severe food insecurity in low-income countries rose by 6.7 percentage points and severe food insecurity by 3.5 percentage points, coinciding with sharp food-price inflation, which is traditionally associated with greater prevalence of acute malnutrition in children \citep{fao2025}. Projections indicate that a 20\% global reduction in N, P, and K fertilizer supply for one year would raise the Food and Agriculture Organization's (FAO) Food Price Index by up to approximately 6\% between 2025 and 2028, and that two consecutive years of 20\% supply shocks would push the index by around 13\% \citep{oecd2024}.

To provide insights into the robustness or fragility of the globally integrated food supply, several studies have addressed how individual parts of the energy--fertilizer--food production and trade nexus may be affected by disruptions. \citet{ahvo2023} quantify how fertilizer and other agronomic input shocks affect production, finding that losses concentrate in high-yield regions; however, they treat input shocks as exogenous, globally uniform losses and do not model where shocks originate or how they propagate through trade. \citet{li2024} investigate how shocks affect the trade and availability of selected staple crops over time within the international trade network. Accounting for the international distribution of food production, processing, and consumption, \citet{laber2023} trace shock propagation through a multilayer food trade and processing network, demonstrating that upstream shocks in raw commodities can have larger downstream effects than shocks to final food products themselves. Concerning agronomic input supply chains, \citet{li2023} evaluate the stability of fertilizer trade networks. Cascading supply chain disruptions and commodity production impacts have been demonstrated in a few conceptual studies for non-food sectors \citep{lee2016, wei2022, kang2024}. Yet this approach has not been pursued for the energy--fertilizer--food chain, despite its critical relevance for global food security.

\begin{figure}[h]
    \centering    \includegraphics[width=1\textwidth]{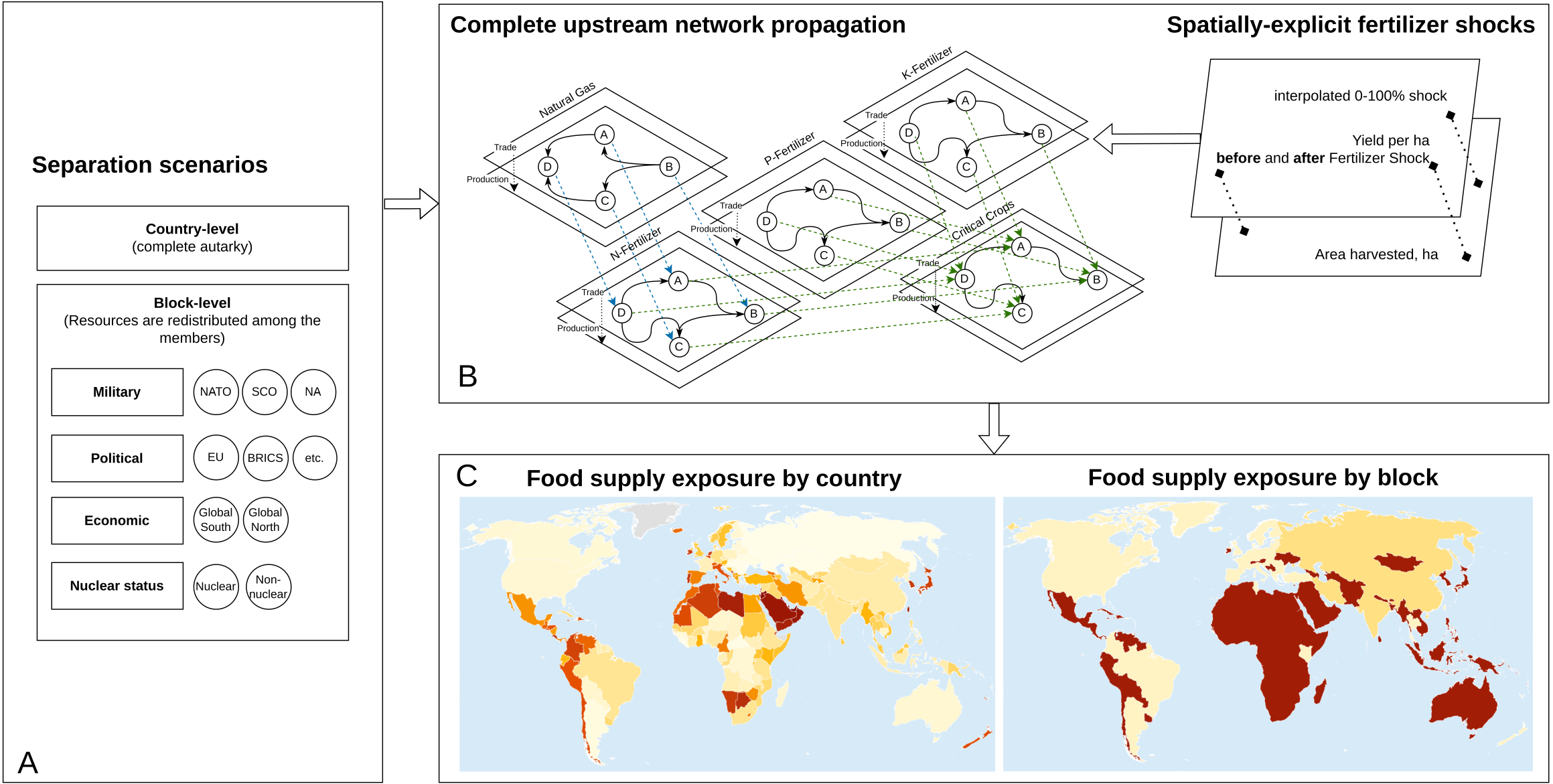}
    \caption{\textbf{Schematic of the study design.} (A) Scenarios define disruptions within the trade network occurring at the country level, for specific types of country blocs or alliances. (B) Trade flow disruptions, resulting fertilizer and crop production losses, and ultimately national or bloc food supply are modelled by combining a chain of network and impact response models. (C) Potential impacts on food supply are evaluated at the country or bloc level, combined with panel data on socio-economic characteristics. Details are provided in the Methods section, an overview of bloc configurations and their rationales in \autoref{tab:scenarios})}
    \label{fig:cascade}
\end{figure}

Here we employ a global trade and impact cascade model (\autoref{fig:cascade}) to quantify food security vulnerabilities to agronomic and food trade disruptions for eleven critical staple crops across all countries globally and twenty bloc configurations organized into economic, political, military associations, nuclear weapon-bearing, and non-aligned countries. The country and bloc configurations consider that in the case of major trade disruptions that can have economic, military, or political causes, countries may decrease or cease bilateral trade \citep{long2008} but may also reinforce integration into blocs consistent with current geopolitical trends \citep{gopinath2024}.We simultaneously model sequential disruptions in natural gas trade, the key pre-cursor for nitrogen fertilizer production, trade in macro-nutrients N, P, and K fertilizers which in turn reduces crop production capacity, and eventually trade in crop commodities. Based on spatially explicit shock responses, we estimate potential production losses from trade disruptions at each cascade stage and aggregate results by country and defined blocs. To provide insights on market powers and their concentration, we furthermore trace the dynamics of centrality evolution across networks and eventually quantify food stock duration capacities as a means of resilience to disruptions.

\section{Results}\label{results}

\subsection{Food supply vulnerabilities of contrasting country blocs}
\label{sec:blocs}

\begin{figure}[h]
    \centering
    \includegraphics[width=1\textwidth]{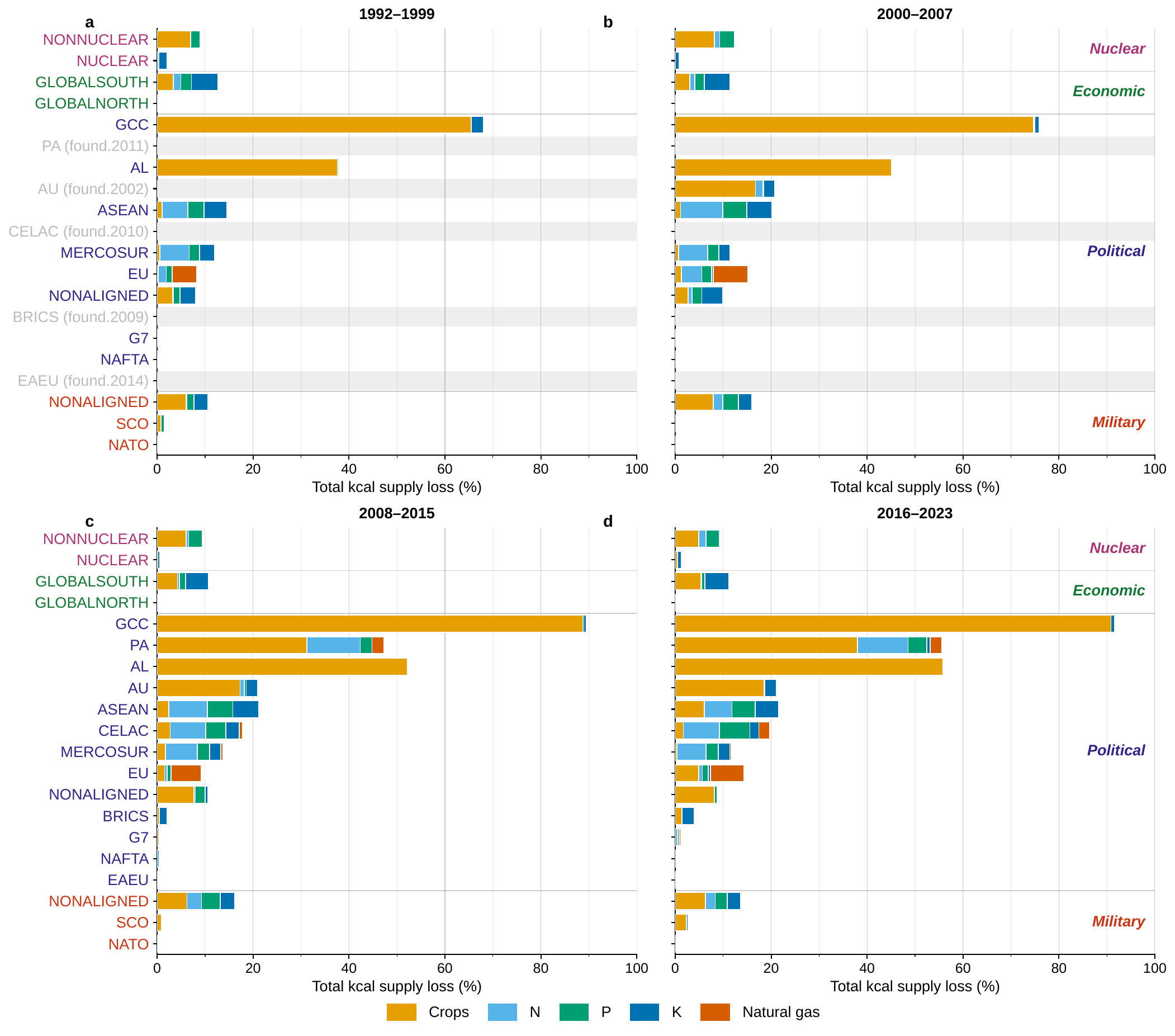}
    \hfill
    \caption{\textbf{Food supply loss by impact driver in twenty geopolitical blocs.} Blocs are grouped into four archetypes: nuclear, economic, political, and military. Bars show the additive contributions of five shock components to total kcal supply loss under trade isolation: crop self-sufficiency failure (Crops), compound disruptions in nitrogen (N), phosphate (P), and potassium (K) fertilizer supply chains, and natural gas propagation losses. Panels (a)--(d) correspond to 1992--1999, 2000--2007, 2008--2015, and 2016--2023, respectively. Blocs not yet founded as of the panel date are shown in muted colors and indicate their foundation date. Data for SCO are available from 1996 onward, corresponding to the foundation of its predecessor, the Shanghai Five.}
    \label{fig:blocks}
\end{figure}

Assuming isolation from the rest of the world, pronounced disparities emerge across blocs in agronomic supply chain dependency and vulnerability (\autoref{fig:blocks}). During the most recent years analyzed (\autoref{fig:blocks}d), simulated trade disruptions project regional crop availability losses ranging from near zero to more than 90\%. Three blocks occupy a distinct \emph{high-shock tier}: the Gulf Cooperation Council (GCC, mean 92\%), the Arab League (AL, 56\%), and the Pacific Alliance (PA, 56\%). The remaining blocks fall into a \emph{mid-tier} including the Association of Southeast Asian Nations (ASEAN, 21\%), African Union (AU, 21\%), Community of Latin American and Caribbean States (CELAC, 20\%), the European Union (EU, 14\%), and militarily non-aligned countries (14\%), a \emph{lower tier} composed of the Southern Common Market (MERCOSUR, 12\%), the Global South (11\%), and non-nuclear-weapon-bearing and politically non-aligned countries (9\%), and a \emph{near-zero tier} with the Eurasian Economic Union (EAEU), Global North, North American Free Trade Agreement (NAFTA), and North Atlantic Treaty Organization (NATO), all below 0.5\%. Four geopolitically prominent blocks sit between the lower and near-zero tiers with mean losses of 1--4\%, the BRICS (4\%), Shanghai Cooperation Organization (SCO, 3\%), nuclear-armed states (1\%), and G7 (1\%). Their modest collective exposure reflects the structural food and fertilizer self-sufficiency of the world's leading economies.

The assessment also reveals structurally distinct shock profiles. \textit{Crop-dominated} (crop self-sufficiency failure is the primary driver) are GCC, AL, PA, AU, non-nuclear, and both non-aligned blocs. GCC and AL have nearly sole crop trade-driven vulnerabilities ($\geq$99\% crop imports fragility); PA is mixed, with the crop component the largest driver (38\%), followed by N fertilizer (11\%) and natural gas (2\%). The \textit{N-dominated} profile includes CELAC, MERCOSUR, NAFTA, NATO, and the Global North, where fertilizer supply chains are the binding constraint. CELAC shows an increasing fertilizer vulnerability trend. ASEAN, by contrast, has shifted into a balanced crop-fertilizer profile: its crop component has increased six-fold over the study period (from 1\% to 6\%) and now marginally exceeds its nitrogen component. The EU is uniquely \textit{natural gas-dominated}. Natural gas propagation losses here are the largest shock component (7\% in 2016--2023), reflecting deep structural dependence on gas-based fertilizer synthesis and energy systems. Beyond this one, G7, CELAC, PA, and MERCOSUR also exhibit non-zero natural gas components, though all are relatively marginal. \textit{Potassium-dominated} profile constitutes of BRICS and nuclear weapon-bearing countries where K supply concentration drives risk; the Global South sits at the crop--potassium margin, with its crop and potassium components nearly tied.

Among all blocs, nuclear-armed countries and the Global South are the only ones exhibiting pronounced declining trajectories of shock vulnerability over the study period (\autoref{fig:blocks}a-d). Most of the remaining blocs have become more vulnerable, while the African Union, militarily non-aligned, non-nuclear, and EAEU blocs show no pronounced trend.

Politically non-aligned countries show an abrupt increase in autarky shock potential from 3\% (1992--2007) to 8\% (2008--2023), more than doubling its pre-break baseline and remaining elevated through the latest period. The GCC exhibits the largest escalation in absolute terms, with its crop imports-driven mean total loss rising monotonically across all four periods (from 68\% in 1992--1999 to 76\% in 2000--2007, 90\% in 2008--2015, and 92\% in 2016--2023), reflecting persistent deterioration in domestic crop self-sufficiency. The Arab League follows a parallel entirely crop-driven trajectory (from 38\% to 45\%, 52\%, and 56\%), with zero fertilizer or natural gas contribution in any period. The EU, for its part, has remained natural-gas-dominated across all four periods with natural gas continuously accounting for the largest single share. Its profile shifted within the fertilizer block (notably a sharp drop in N exposure after 2007) and a marked crop-autarky component has emerged in the most recent period (rising from 0.2\% in 1992--1999 to 4.9\% in 2016--2023). Overall, the EU's total vulnerability is now almost two times higher than in 1992--1999 (from 8\% to 14\%).


\begin{figure}[h]
    \centering
    \includegraphics[width=0.9\textwidth]{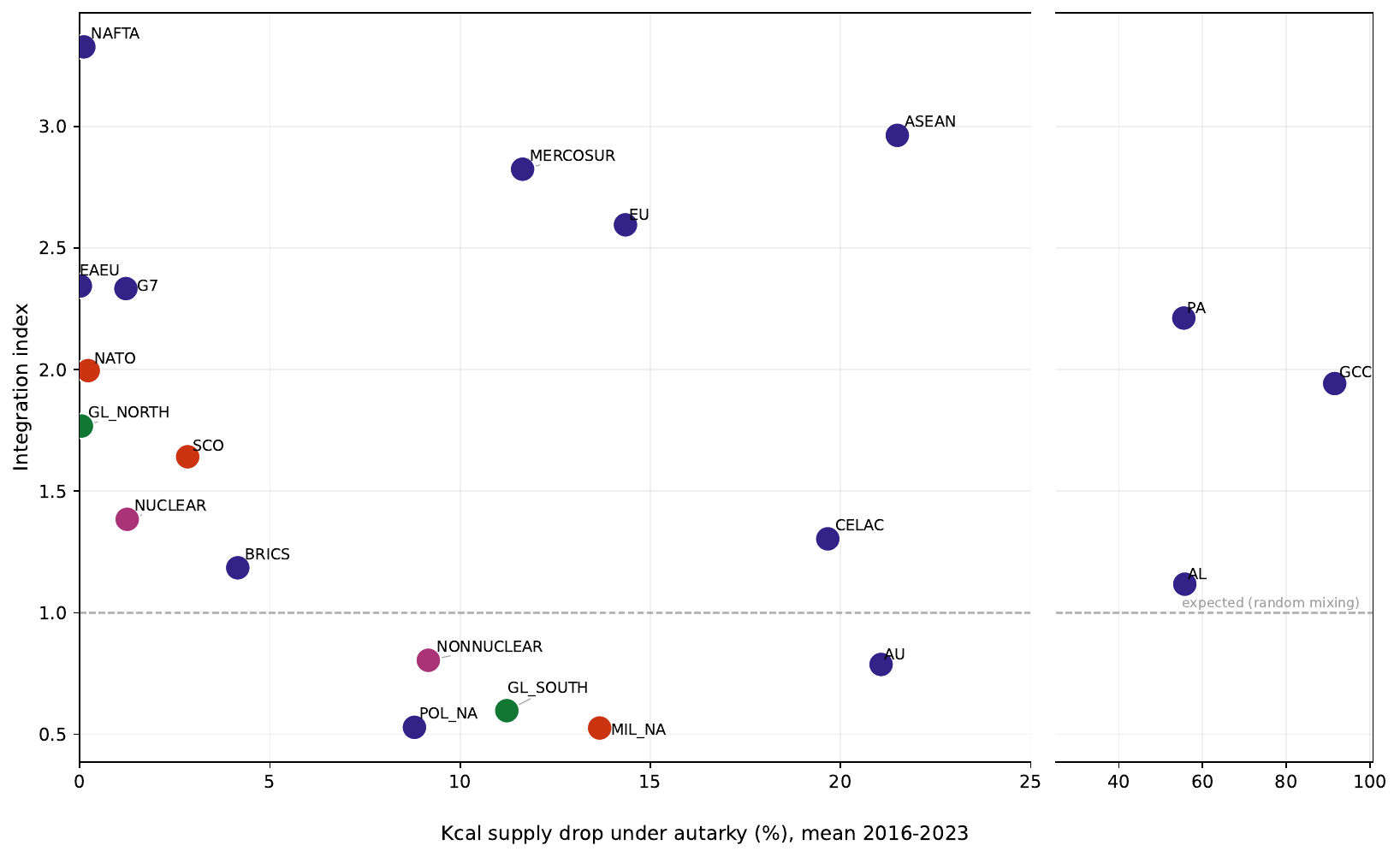}
    \hfill
    \caption{Loss of calorie supply in twenty major country blocs compared to blocs' economic integration. Loss of calorie supply (x-axis) is the same as in \autoref{fig:blocks}d, reflecting the mean over the period 2016--2023. The axis is split at 25\% and the 0--25\% segment is horizontally stretched for better readability. The economic integration index $\mathcal{I}$ (y-axis) measures the degree to which a bloc's aggregate trade is concentrated among its own members, corrected for the mechanical effect of bloc size (see Supplemental methods). The reference value $\mathcal{I} = 1$ corresponds to the null expectation of size-proportional, non-preferential mixing. Values $\mathcal{I} > 1$ indicate that members trade with one another more intensively than bloc size alone would predict -- a signature of internal cohesion generated by common markets, customs unions, shared currencies, or preferential agreements -- whereas $\mathcal{I} < 1$ indicates that members trade disproportionately with outside partners, so that the bloc offers less buffering against external trade disruptions. Colors of points correspond to the scenario colors in \autoref{fig:blocks}: purple for blocs in the political scenario, green for the economic scenario, red for the military scenario, and bordeaux for nuclear and non-nuclear blocs.}
    \label{fig:tradeintegration}
\end{figure}

We also evaluate the economic integration index against the expected food supply loss under bloc separation to assess the potential resilience to disruptions within blocs by sustaining internal trade in the event of isolation, assuming that the existing trade infrastructure remains intact (\autoref{fig:tradeintegration}). The six most economically integrated blocs are NAFTA, ASEAN, MERCOSUR, EU, EAEU, and G7. Moderately integrated blocs are NATO, the Global North, PA, GCC, SCO, nuclear weapon-bearing states, AL, CELAC, and BRICS. The least integrated are non-nuclear-weapon-bearing states, AU, the Global South, and militarily and politically non-aligned countries, all of which -- except for the AU -- present loose groupings of countries. Notably, NAFTA combines the highest integration with near-zero expected loss, whereas the GCC couples the largest expected loss with only moderate integration -- internal cohesion can redistribute supply but cannot substitute for resources a bloc's members collectively lack. In an escalation of political tensions or a major conflict, non-aligned and non-nuclear-weapon-bearing states, as well as the countries of the broader Global South, would accordingly likely struggle to support their populations in relocating resources to where they are most needed, leaving individual countries to fend for themselves. Moreover, high trade integration within blocs does not, by itself, guarantee that member states are necessarily able to assist each other, which is for example the case if trade infrastructures are impaired or physical choke points are blocked. Against this backdrop, we  examine dynamics of trade isolation at the country level.



\subsection{Potential impacts on countries' food security}
\label{sec:countries}

\begin{figure}[h]
    \centering    \includegraphics[width=0.8\textwidth]{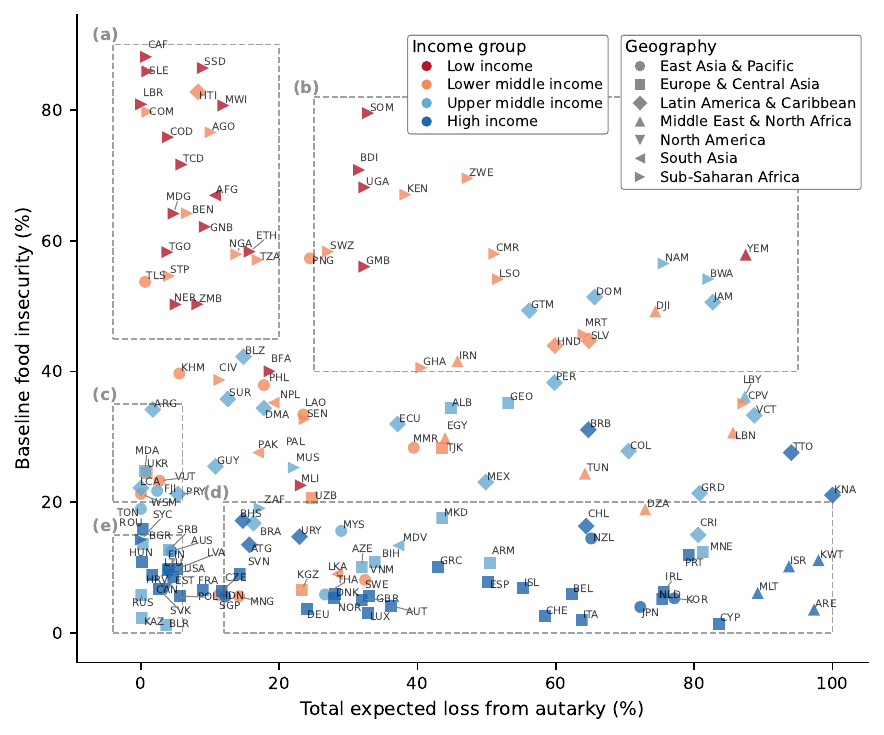}
    \caption{\textbf{Loss of calorie supply in individual countries compared to their baseline food insecurity in the period 2016-2023.} Loss of calorie supply assumes a complete disruption in agronomic inputs and crop trade. The baseline indicator is the FAO's prevalence of moderate or severe food insecurity index adopted from \citet{faostat2024}. Symbols' geometric form indicates the major geographic regions where countries are located, colors distinguish income groups according to the World Bank classification for the same time period \citep{fantom2016}.  The figure displays the 154 countries for which both loss estimates and food insecurity data are available. Dashed boxes labeled a-e indicate visually identified country groupings:  (a) extremely insecure with low expected losses; (b) food insecure and fragile; (c) self-sufficient but food insecure countries; (d) food secure but highly fragile; (e) food secure and resilient.}
    \label{fig:undernourish}
\end{figure}

A scenario of complete global autarky unravels individual countries' food supply fragility in cases of sole self-reliance. Comparing food insecurity indices against potential total losses of calorie supply resulting from cascading trade and production disruptions, provides two key insights regarding countries' economic status and geographies (\autoref{fig:undernourish}).

First, high and upper middle income countries (\autoref{fig:undernourish}; symbols colored dark blue and light blue) are not immune to losses in critical crops. While they are presently food secure they scatter across the entire range of potential losses in food availability, manifesting fragility with up to 100 percent decrease in the availability of critical calories supply if trade is disrupted. Second, low-income countries (\autoref{fig:undernourish}; symbols colored red) exhibit higher rates of food insecurity, yet on the x-axis they scatter only up to 30--35 percent (with Yemen as an outlier at 87\%). This pattern may intuitively suggest greater robustness of low-income countries. However, given that the food insecurity situation in these countries is already dire, low decreases in food supply under disruption is a symptom of already prevalent scarcity rather than robustness. In addition, these seemingly smaller consequences will affect already food insecure populations disproportionately.

Geographically, five country categories emerge (\autoref{fig:undernourish}): (a) \textit{extremely food insecure with low expected losses} including foremost landlocked, low productive Sub-Saharan countries such as Chad, Niger, Democratic Republic of Congo, Madagascar, Liberia, Sierra Leone, or Central African Republic); (b) \textit{food insecure and fragile} with Central America and fertilizer supply chain-dependent West and Central African states such as Ghana, Cameroon, Gambia, Uganda, and Burkina Faso, where fertilizer exposure substantially exceeds crop-autarky risk; (c) \textit{self-sufficient but marked as food insecure middle-income countries} such as the Republic of Moldova, Ukraine, and Argentina; (d) \textit{food secure yet highly fragile} with Persian Gulf, Levante, and North African states, middle to high income East Asian countries (Japan, South Korea, Taiwan, Hong Kong), Western and Mediterranean European states, and most small-island economies, with losses predominantly crop-autarky-driven;  and (e) \textit{food secure and resilient}, mainly the Eurasian grain-export cluster Russia, Belarus, and Kazakhstan, Central and Eastern European key grain producers such as Hungary, Romania, Bulgaria, and Croatia, the North American exporters Canada and USA, and Australia. Across the 193 countries, crop-autarky is the primary loss driver for $\approx$122, N for 22 countries, most extreme in parts of Northern Europe, K for 15 countries, and natural gas is the dominant single component in Germany and Poland.





\subsection{Impacts on key crop and fertilizer net-exporters}
\label{sec:exporters}

\begin{figure}[h]
    \centering
    \includegraphics[width=1\textwidth]{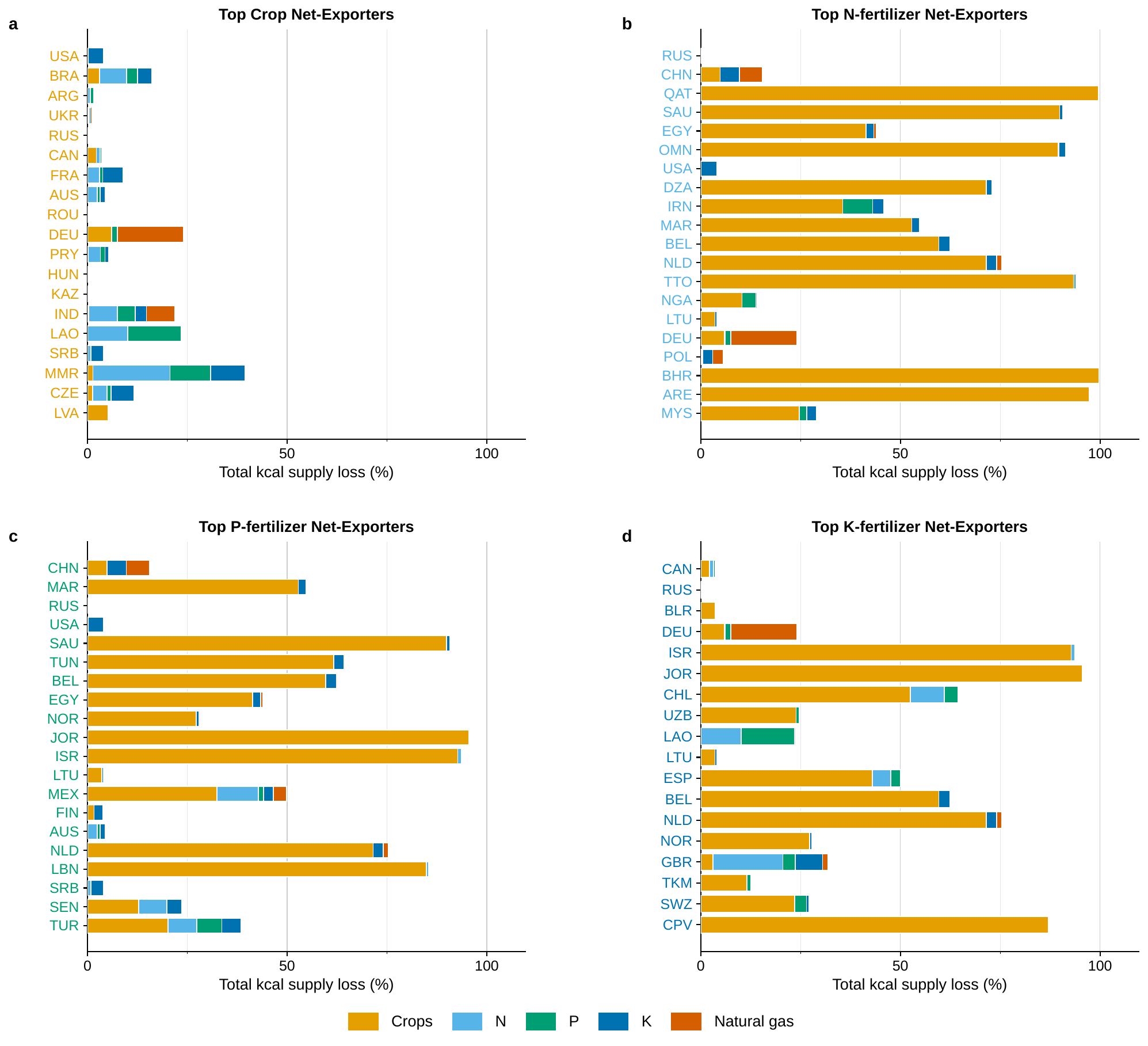}
    \hfill
    \caption{\textbf{Vulnerabilities of twenty top net exporters for crops and fertilizers to loss of calorie supply in the period 2016-2023.} Each bar indicates total losses in food calorie supply due to disruptions in supply across three networks covering crops, fertilizer (three macro nutrients), and natural gas. N=nitrogen fertilizer, P=phosphate fertilizer, K=potassium fertilizer. Panels show order by total net export volume key (a) crop, (b) N fertilizer, (c) P fertilizer, and (d) K fertilizer exporters. Some countries appear in several panels if they are key exporters for several of the commodities.}
    \label{fig:country}
\end{figure}

To provide insights into the vulnerabilities of countries of strategic importance within supply chains, we evaluate outcomes for major net-exporters of crops and fertilizers (\autoref{fig:country}). Overall, crop exporters are more resilient to supply chain disruptions than fertilizer exporters, and a narrow cluster of net-exporters is structurally most robust under autarky, i.e. the Eurasian grain belt including Russia (0.06\%), Kazakhstan (0.1\%), Hungary (0.16\%), Romania (0.2\%), and Ukraine (0.8\%) together with Argentina (1.7\%), Canada (3.6\%), the USA (4\%), and Australia (4.4\%), all of which retain more than 95\% of caloric supply across all three supply chain parts (\autoref{fig:country}a). These countries are characterized not only by disposing of sufficient land and suitable climate for crop production, but also by resource endowments covering natural gas and crop nutrients, at least to some degree. The most vulnerable net crop exporters are predominantly located in Southern Asia (India, Laos, Myanmar) but also include Brazil, a crucial breadbasket, and Germany. The latter is notably dependent on natural gas imports, while the aforementioned countries face constraints across several crop nutrients.

Conversely, most fertilizer exporters are highly dependent on crop import. Twelve of the top twenty N fertilizer exporters are prone to lose over 40\% of their caloric supply under autarky, and six -- namely, Qatar (99.6\%), Bahrain (99.7\%), the UAE (97.3\%), Trinidad and Tobago (94\%), Oman (91\%), and Saudi Arabia (91\%) would face near-total collapse (\autoref{fig:country}b). In P fertilizer trade, Morocco, which holds more than two thirds of known global phosphate-rock reserves, sources more than 50\% of calorie supply from crop imports as do several other countries from the MENA region, namely Jordan (95.6\%), Israel (93.7\%), Tunisia (64.2\%), and Lebanon (85.6\%). Turkey, although it is a P fertilizer net-exporter, has a notable exposure to P fragility (6.3\%) (\autoref{fig:country}c). The United Kingdom ranks among the top K-fertilizer net-exporters, yet, its own loss profile contains K fertilizer as a driver on top of a dominating N import dependency (N 17.4\%, K 6.8\%) (\autoref{fig:country}d). Germany (24\%, natural gas 16.5\%), Poland (2.6\% of 5.6\%), and India (7.1\% of 22\%) are the only major exporters with notable natural gas loss components. Southeast Asian crop exporters -- Myanmar (39.4\%), Laos (23.5\%), and India (21.9\%, with fertilizer channels accounting for 66\% and natural gas for a further 33\% of the loss) -- exhibit markedly higher vulnerability than their temperate-zone counterparts, with fertilizer channels rather than crop trade carrying most of the loss.

\subsection{Evolution of market power concentration in trade networks}
\label{sec:centrality}

\begin{figure}[H]
    \centering
    \includegraphics[width=0.9\textwidth]{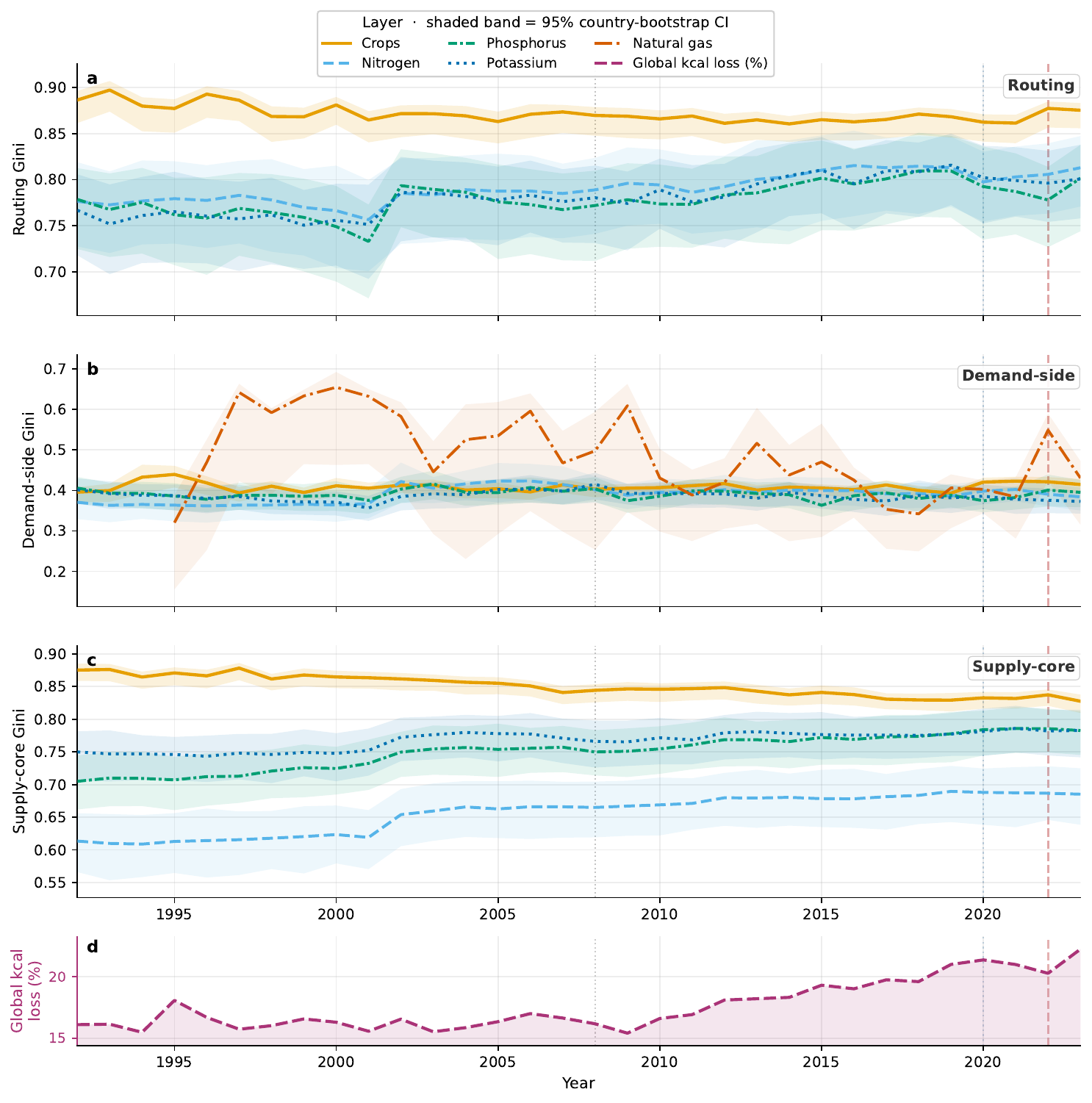}
    \hfill
    \caption{\textbf{Concentration of market power across commodity layers, 1992--2023.} \textbf{Top three panels:} betweenness (routing), PageRank (demand-side) and eigenvector (supply-core) centrality. \textbf{Bottom panel:} global compounded caloric loss under the autarky shock model. Crops aggregate 11 constituent layers; nitrogen (N), phosphorus (P), potassium (K) and natural gas are plotted as single layers. Natural gas is shown only in the demand-side (PageRank) panel and is omitted from the routing and supply-core panels, where its almost strictly bipartite trade structure leaves no transit role and a degenerate single-block core. Betweenness centrality identifies countries that lie on shortest trade paths between others, i.e. transit and brokerage hubs whose disruption fragments the network. PageRank captures recursive influence, in which a country's importance increases with the importance of its trading partners. Eigenvector centrality measures membership in the densely interconnected core of mutually reinforcing trade relationships. We summarized the country-level distribution of each component, within each layer-year, using the bias-corrected Gini coefficient. Values closer to $0$ indicate that all countries are relatively equally central, values closer to $1$ are an indicator that few countries dominate.}
    \label{fig:centrality}
\end{figure}

Different centrality concentration measures can be applied to resolve where market powers and, hence, fragility are located in the global energy-fertilizer-food supply chain. Here, we measure what we term (I) \textit{routing} concentration as a measure of trade path length that identifies importance of transit and hubs, (II) \textit{demand-side} concentration as a metric of how important countries are for final destination of goods, and (III) the proximity to densely clustered \textit{supply-core}, in other words, membership in a mutually reinforcing, densely interconnected export bloc (\autoref{fig:centrality}).

%
Crop trade is more concentrated in routing and supply-core centrality than fertilizer trade, reflecting the hub-and-spoke structure of staple-grain trade in which a small set of large exporters routes the bulk of global flows (\autoref{fig:centrality}a-c). N shows the most distributed supply-core profile, consistent with the geographic breadth of ammonia synthesis from natural gas, whereas P and K are substantially more concentrated, resulting from the geological concentration of phosphate reserves in Morocco and China and of potash in Canada, Belarus, and Russia. Demand-side metrics suggest that all crop trade dynamics present similar moderate concentration, indicating that recursive trading influence is shared across a substantially broader set of participants than transit routing or supply-core membership. Routing and supply-core concentration are hardly detectable for natural gas due to the almost strictly bipartite gas trade, meaning that one or a few exporters predominantly ship directly to many importers. On the demand side, where natural gas can be compared with the other layers, its concentration is both elevated and by far the most volatile: its top-5 importer share far exceeds that of crops and individual fertilizers (\ref{fig:topk}).

Long-term trends in market concentration are modest in crops but pronounced in fertilizers (\autoref{fig:centrality}a--c). The clearest feature is a step increase in fertilizer concentration around 2001--2002, visible across all three measures and all three nutrients yet absent in crops: supply-core concentration steps from $0.6$ (pre-2002) to $0.7$ (post-2002) for nitrogen and from $0.7$ to $0.8$ for phosphorus, with parallel upward shifts in routing and demand-side concentration. Thereafter, the upstream and downstream layers diverge: fertilizer supply-core concentration continues to drift upward, whereas crop supply-core concentration declines steadily, and crop demand-side concentration remains broadly flat. Natural gas demand-side concentration is the least stable, ranging between $0.3$ and $0.7$ over the period, with its sharpest single-year jump in 2022 (Gini from $0.4$ to $0.6$) amid the European gas crisis following the war in Ukraine.


These structural features shift the locus of fragility upstream. Fertilizer trade has grown steadily more concentrated since the early 2000s, while natural gas -- the feedstock for nitrogen synthesis -- is the most volatile and least diversified layer, and a structurally distinct, infrastructure-bound system (\autoref{app:netcomms}); crop trade, by contrast, remains comparatively stable across all three measures. Because concentration narrows the set of substitutable suppliers, this configuration could potentially propagate disruptions originating in the gas and fertilizer layers into downstream crop calorie supply, as quantified by our shock-propagation model.

The global compounded caloric loss under full autarky shows a structural break around the 2008 financial crisis (\autoref{fig:centrality}d). From there onward the prior flat slope turns into an upward trend with potential loss increasing at $+4$ percentage points per decade, reaching an all-time high of $22\%$ in 2023, which is also the year with the largest single increase ($+2$ pp from 2022 to 2023). The combination of high-volatility upstream layers and a steadily rising autarky cost is consistent with a system becoming increasingly sensitive to disruptions in precisely those upstream layers where market power is most concentrated and volatile.

Further discussion of the role of individual countries and their profiles is provided in \autoref{app:centrality}. A complementary community-detection analysis (\autoref{app:netcomms}) confirms this architecture: crop and fertilizer trade place countries into the same structural roles -- a single integrated agri-food complex -- whereas natural gas is structurally coupled to the agri-food system only through gas-dependent nitrogen synthesis, the structural signature of the upstream chokepoint.

\subsection{Capacity of food stocks and domestic production}
\label{sec:stocks}

Domestic agronomic input and food stocks can buffer supply shocks resulting from trade and production disruptions \citep{marchand2016}. However, their national volumes are not systematically recorded and are frequently part of countries’ strategic contingency plans, rendering their amount subject to uncertainties. In recent years, the FAO has begun to estimate the size of annual crop commodity stocks \cite{faostat2024}. Therefore, we do not include stocks in the modelling of trade disruptions directly but instead evaluate complementary how long reported stocks or domestic production would support countries' present crop commodity consumption levels.

\begin{figure}[H]
    \centering
    \includegraphics[width=1\linewidth]{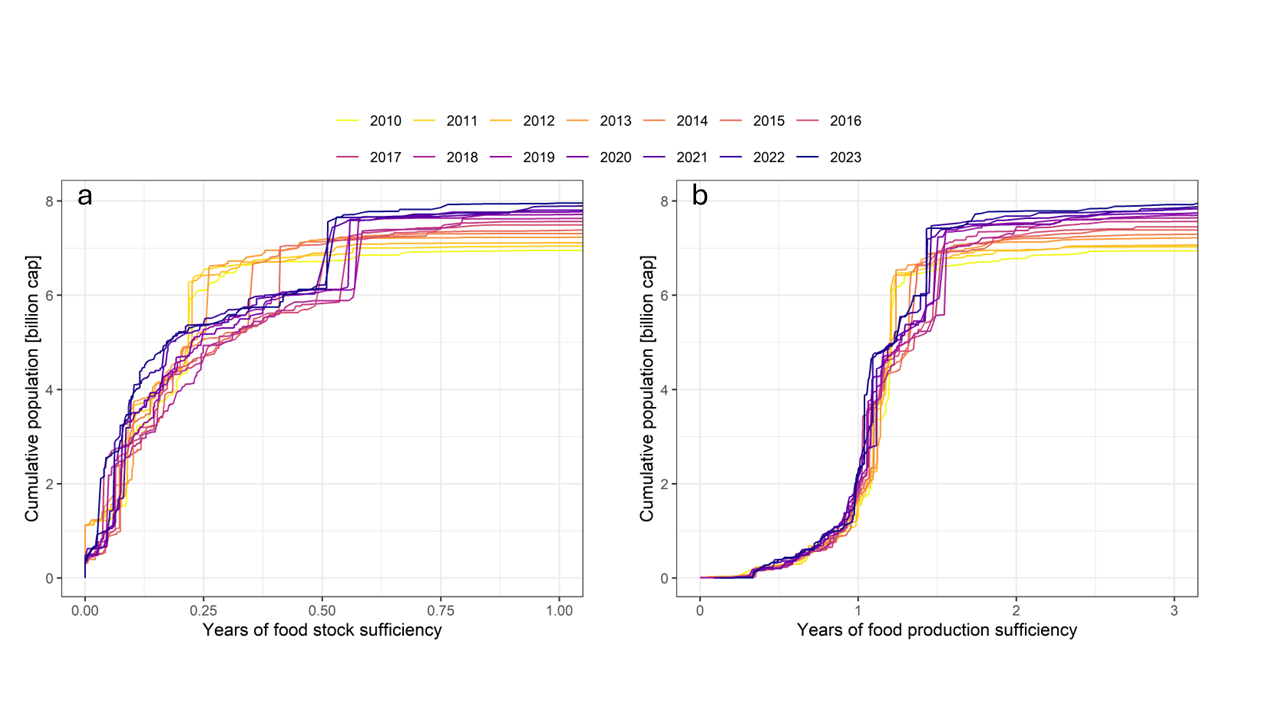}
    \caption{\textbf{Global populations by duration of domestic food stock and production.} Panels show cumulative population numbers ordered by the duration of (a) domestic food stocks and (b) domestic food production to maintain present consumption levels for the eleven key crops considered herein in the time period 2010-2023. Calculations are performed based on calorie equivalents of both supply and consumption. Annual population fractions are shown in \ref{fig:stockdurationrel}.}
    \label{fig:stockduration}
\end{figure}

Almost the entire global population lives in countries with known stocks lasting less than one year, three quarters of the global population in countries with stocks lasting less than half a year, and half the population in countries with stocks lasting less than three months (\autoref{fig:stockduration}a). Stocks can accordingly provide only very short-term buffering. This vulnerability has manifested in severe food shortages in several important-dependent low- to middle-income countries during recent, rapidly evolving global food supply crises \citep{glauber2023, alexander2023}. A country with large population and comparably large stocks is China, visible in the upper quarter of the figure as a line segment moving between 0.2 and 0.6 years. Here, the potential duration of stocks has substantially increased over the observed time period, though a stagnation and slight rebound in most recent years are visible. This trend has also occurred, albeit less pronounced, in many other countries, as indicated by the stacked lines. India, as a comparably populous country, is located at the lower end with food stocks lasting approximately one month (\ref{fig:stockdurationrel}).

Domestic production can last substantially longer -- for approximately half to three quarters of the global population for more than a year, and for a small fraction (<5\%) for more than two years  (\autoref{fig:stockduration}b). Here, too, a marginal trend toward longer-lasting supply over time is apparent, with a recent rebound. Consequently, provided that stocks for the vast majority of countries last less than one year, and domestic production for at least a quarter of the global population less than one year as well, the timing of an agronomic or food supply chain disruption would determine how long food supply can be met from these two sources. Essentially, they provide resilience successively when cultivation is already under way in a given year while stocks are being consumed.

\section{Discussion}\label{discussion}


In essence, our results highlight divergent vulnerabilities among countries and blocs to distinct parts of the supply chain disruptions analyzed herein (see \autoref{sec:blocs} and \autoref{sec:countries}). For example, the Persian Gulf region depends almost exclusively on crop commodity imports, while countries of the Global South frequently rely on imports of crops and potassium fertilizers. The EU has the capacity to produce N fertilizers and play a substantial role in fertilizer exports, but it faces primary vulnerability to natural gas supply disruptions. Latin America, in turn, is critically dependent on N fertilizer imports, while constituting one of the major food crop export regions globally. African nations, finally, are vulnerable to both direct food import disruptions and fertilizer scarcity, while often having large food-insecure populations.

Net exporter status is not equivalent to food self-sufficiency. Only a small cluster of temperate grain producers retains over $95\%$ of caloric consumption under autarky. Most major fertilizer exporters are themselves highly vulnerable, with a substantial share prone to lose more than $40\%$ of their caloric supply and the most extreme cases approaching near-total collapse (see \autoref{sec:exporters}). This fragility typically runs through a different commodity layer than the one in which a country dominates global markets, leaving export rank, dietary self-sufficiency, and shock resilience largely decoupled across the global food trade system.

Market power in the global energy-fertilizer-food trade is unevenly distributed, and the inequalities are most pronounced upstream. A small group of countries dominates trade in natural gas, phosphate and potash, while crop trade relies on a somewhat broader, though still hub-dominated, set of exporters (see \autoref{sec:centrality}). These upstream layers are also the most volatile year-to-year, and natural gas in particular is a structurally distinct, infrastructure-bound system with few substitutable partners; a disruption entering through this narrow, non-substitutable gateway can therefore propagate downstream into crop supply, a structural exposure that our shock-propagation model quantifies. In parallel, the global cost of trade disruption shows a structural break around 2008: the share of calories at risk under a full-autarky shock has been increasing, reaching an all-time high of about $22\%$ in 2023. The global food system has therefore become increasingly sensitive to potential shocks.

This is further complicated by the finding that nearly the entire world population lives in countries whose stocks cover less than a year of consumption; specifically, three quarters reside in countries with less than six months, and half with less than three months (see \autoref{sec:stocks}). The timing of a shock relative to the agricultural calendar largely determines how far domestic resources alone can carry a population, leaving import-dependent low- and middle-income countries that hold the smallest stocks chronically exposed. Critically, almost no country is immune to food security collapse regardless of development status; existing food insecurities will be disproportionately amplified in already vulnerable countries, creating humanitarian emergencies requiring coordinated anticipatory response with long-term consequences for global stability.


Options to mitigate cascading supply chain risks in the short and long term are highly diverse and context-dependent. In the short term, these include strategic stockpiling, diversification of trade partners, and activation of dormant trade links \citep{wood2023, marchand2016}. Over the longer term, options range from developing domestic fertilizer production via green ammonia -- particularly for countries with high renewable energy potential \citep{quitzow2025} -- to strengthening supply chain diversity and regional market connectivity \citep{jia2024, wood2023}. Which options are viable depends critically on countries' environmental endowments (e.g., renewable energy and water availability), economic development level, and geopolitical positioning \citep{quitzow2025, jia2024, schneider2023}. Many of these options entail trade-offs with other goals: decarbonization of fertilizer production may increase short-term costs and reduce affordability for low-income countries \citep{quitzow2025}; trade openness can enhance food availability but erode food sovereignty \citep{wood2023}; and prioritizing domestic self-sufficiency may conflict with efficiency and achieving sustainability goals \citep{jia2024, schneider2023, nystrom2019}.

Since the onset of the conflict in Ukraine, for example, the EU has in part set aside environmental sustainability goals relating to biodiversity, climate, and soil health enshrined in the Farm to Fork strategy to prioritize boosting domestic food production \citep{mangnus2025}. Yet it remains dependent on the gas and fertilizer imports demonstrated herein and is therefore vulnerable to supply chain disruptions. Critics of this policy reversal have accordingly stressed that a food system transformation less reliant on exogenous inputs remains the most sustainable option, both with respect to existing sustainability goals and with respect to increasing food supply resilience \citep{mangnus2025, portner2022}. In fact, the EU and other regions have already initiated cross-sectoral efforts fostering improved plant nutrient management and nutrient circularity -- thus far especially for P \citep{walsh2023}.

Some major fertilizer producers and consumers, such as China, have in turn introduced export restrictions on fertilizers (and food) and have in part stepped up domestic production in recent years in response to global supply chain crises beginning with the 2020 epidemic \citep{glauber2023, vos2025}. Addressing not only supply chain resilience directly but also the need to decarbonize fertilizer production, electrification is emerging as an alternative pathway for decentralized N fertilizer production \citep{rosa2023} that can in principle also be rolled out in presently resource-constrained regions with potential for non-fossil energy production \citep{tonelli2024}. If scalable, this pathway would simultaneously allow for fertilizer and food sovereignty where other resources, such as land and water, are sufficiently available.

As a strategy immediately relating to supply chain resilience, countries are increasingly returning to strategic stockpiling of both food and fertilizer resources \citep{amaglobeli_policy_2023, oecd2024}. Such stocks are often subject to concealment and are therefore largely unknown. For food supply, we estimate herein stock duration and find that most people globally are highly vulnerable to shocks lasting more than a few months. Finally, contrasting the present drive toward increasing fragmentation among countries, our results underpin that countries are generally better off participating in alliances where complementary exchange of goods along supply chains persists and stockpiles can be shared among members in cases of emergency, in line with recent literature \citep{kuhla2024}. Hedging strategies, as observed during recent food supply chain crises \citep{glauber2023, vos2025}, therefore appear unlikely to constitute a sustainable solution.


While based on the most recent data and state-of-the-art modelling approaches, our study is subject to several assumptions and limitations summarized here (see \autoref{app:limits} for more detailed elaborations).  First, we assume that any loss in resources for fertilizer or crop production affects all products proportionally and uniformly. In practice, this is not to be expected, as comparative advantages and policy interventions will alter the distribution of impacts. Yet, in the context of our study this approach facilitates transparent interpretability of outcomes, provided this limitation is borne in mind. Furthermore, as outlined previously, gas, fertilizer, and crop stocks are often unknown and can support countries' coping abilities. Such buffers will, however, result only in limited delays before impacts materialize. Finally, we do not account for transport routes and trade infrastructure, such as ports. While their relevance has been quantified from an economic perspective for maritime transport  \citep{verschuur2025}, more granular data on commodity flows along transport routes and in time will be required to incorporate this dimension into our modelling approach.

\section{Concluding remarks}\label{conclusion}



For the majority of countries, maintaining open trade routes is a food-security necessity. The trajectory of openness is, however, no longer the default planning assumption. As the world drifts toward fragmentation along bloc lines and the consolidation of regional macro-economies in most recent years \citep{zahoor2023,gopinath2024,airaudo2025}, the second-best policy is to deepen complementarity within one's own bloc so that intra-bloc flows can absorb a shock to extra-bloc trade. Where openness cannot be guaranteed, sovereignty over critical inputs and intra-bloc complementarity become the necessity of resilience policy.

This fallback, however, works only for those who sufficiently integrate with other countries. The politically and militarily non-aligned states and low- to middle income countries in the broader Global South already face the world's highest baseline food-insecurity rates, hold the smallest reserves, and trade least among each other. In any scenario in which the major economic and military blocs reduce or seize their external trade, these countries have neither external partners to redistribute resources to them nor sufficient common markets to do so themselves. As the nearly 80-year period following the Second World War, commonly termed the `long peace,' is statistically insufficient to declare a decline in human conflict propensity  \citep{cirillo2016a,clauset2018}, the risk of major future conflict persists, and our analysis shows that non-aligned countries would presently be the most vulnerable. Consequently, a multilateral architecture for the gas--fertilizer--food nexus that maximizes resilience for the global population appears the most promising strategy to mitigate risks from trade disruptions.

\section{Methodology}
\label{sec:methods}
\subsection{Study design}

The study's overarching objective is to quantify impacts of disruptions in international agronomic supply chains -- encompassing natural gas, fertilizers, and crop trade -- on the supply of 11 key staple crops and, ultimately, food security. As sketched in \autoref{fig:cascade}, we combine network modelling for the representation of trade flows with impact modelling to estimate production losses caused by decreased availability of resources. That is, we estimate the impact of natural gas supply reductions on N fertilizer production and the impacts of decreased fertilizer availability on crop production. These production losses are subsequently forward through the next layer of the trade network.

Two parallel but structurally distinct pipelines were implemented. The first (\emph{country-level}) computed autarky shocks and fertilizer-mediated production losses for 208 countries individually. The second (\emph{bloc-level}, see bloc information in \autoref{tab:scenarios}) pooled production and supply across bloc members \emph{before} computing autarky shocks, and applied bloc-specific crop-response coefficients that account for changing membership composition over time. This dual design yields complementary perspectives: the country-level pipeline preserves national heterogeneity, while the bloc-level pipeline captures the collective exposure of blocs that may pool resources internally.

We produce annual outcomes for temporally transient evaluations of network and impact evolution over the period 1992--2023, as well as averages for selected eight-year periods. The latter smooth inter-annual variations allow for more detailed evaluations at the country and regional level, comprising the periods 1992--1999, 2000--2007, 2008--2015, and 2016--2023. The subsequent sections detail the individual modelling steps and data sources.
{
  \renewcommand{\arraystretch}{1.2}
\begin{table}[htbp]
\centering
\begin{tabular}{p{1.5cm} p{4.5cm} p{5.5cm}}
\hline
\textbf{Scenario} & \textbf{Blocks} & \textbf{Rationale} \\
\hline
Military       & The North Atlantic Treaty Organization (NATO), The Shanghai Cooperation Organisation (SCO), Non-aligned. & This scenario is based on the major military bloc configuration. Although SCO is not a military block per se, it has strong anti-terrorism coordination mechanism and the member states conduct regular joint military exercises. It is common in specialized academic literature to compare military powers of these two formations \citep{sun2018}. \\
Economic       & Global North, Global South. & This separation is based on the development and economic status defined by the UN Conference on Trade and Development (UNCTAD). The scenario helps test the hypothesis about a major conflict outbreak in one of the blocks with the following complete shutdown of trade between these two parts of the world, e.g. a nuclear exchange leads to a total devastation in the Global North, and we estimate whether Global South countries would still be able to sustain themselves (taking into account only immediate trade effects, not nuclear weapon explosion consequences). \\
Political      & The Association of Southeast Asian Nations (ASEAN), the European Union (EU), The Group of Seven (G7), The Southern Common Market (MERCOSUR), Community of Latin American and Caribbean States (CELAC), Pacific Alliance (PA), North American Free Trade Agreement (NAFTA)\footnote{the United States–Mexico–Canada Agreement (USMCA) as of 2020}, the League of Arab States (AL), The Cooperation Council for the Arab States of the Gulf (GCC), BRICS, the Eurasian Economic Union (EAEU), The African Union (AU), Non-aligned.  & This scenario is based on participation in major international associations and agreements. The objective of the particular selection includes to maximize the coverage of the whole world with minimum overlaps. The list includes political, political-economic, purely economic alliances, and the countries participating in none of the above united into one non-aligned bloc. We acknowledge that not all countries within these blocs may choose to cooperate among themselves in case of major perturbations. \\
Nuclear status   & Nuclear-weapon bearing countries, non-nuclear countries. & This scenario does not represent an actual grouping of countries but instead classifies them according to their nuclear power status. It groups nuclear-weapon states, some of which are involved in disputes or armed conflicts, and groups the rest of the world to evaluate the power and dependencies of these country typologies within the food supply chain. This, in turn, can help assess the collective leverage that non-nuclear-weapon states may exert to constrain nuclear-weapon states in potential conflicts through coordinated economic pressure, as well as the extent to which the rest of the world depends on nuclear-weapon states.\\
\hline
\end{tabular}
\caption{Country bloc typologies and their rationales.}
\label{tab:scenarios}
\end{table}
}

\subsection{Natural gas supply effects on nitrogen fertilizer production}

Natural gas is the key source for energy and hydrogen required in the production of N fertilizer through the ammonia pathway of the Haber-Bosch process. We estimate impacts of loss in natural gas on N fertilizer applying the conversion coefficient of 0.65 t NH3/t NG from a global study addressing energy efficiency in fertilizer production \citep{gabrielli2023} and a volume-to-weight conversion coefficient of 0.76 $kg/SCM$ is applied for natural gas (where $SCM=kg/0.76$) \citep{snamatlas2025}, and a coefficient of approximately 735 $kg/m^3$ for liquid natural gas (LNG), derived from the relation 1 Mt LNG $ \approx 1.36$ BCM \citep{BP2021}. Accordingly, N fertilizer production is affected proportionally to natural gas availability, assuming no shifts in the allocation priorities for remaining gas supplies within a given country.

We source gas trade and supply data from two sources: (a) UN Comtrade trade data as provided in a cleaned and harmonized version by CEPII-BACI \citep{gaulier2010, cepii2025} and (b) Energy Insitute's data for countries' energy carrier consumption \citep{ei2023}. Data from UN Comtrade may produce negative values for domestic consumption, which are associated with data quality issues in re-exports that lead to double counting and hub effects \citep{hu2022}. Negative values and values smaller than 0.5 bcm were substituted with values from \citep{ei2023} or set to zero if absent in the database.




\subsection{Fertilizer trade and impacts on crop yield}

Bilateral fertilizer trade data were sourced from the detailed trade matrix of the FAO Statistical Database \cite{faostat2024}. Data are reported therein in the form of the three primary nutrients -- elemental nitrogen (N), phosphate (P\textsubscript{2}O\textsubscript{5}), and potassium oxide (K\textsubscript{2}O -- allowing for direct conversion to elemental nutrient inputs of N, P, and K as used in the crop yield impact modelling below.

Crop yield impacts of changes in fertilizer availability were estimated based on a study and data by Ahvo et al. \citep{ahvo2023, ahvo2023d}, who modelled the impacts of stylized, globally uniform shocks in nutrient supply on 11 key staple crops -- namely barley, cassava, groundnut, maize, potato, rice, sorghum, soybean, sugar beet, sugarcane, and wheat. The authors provide gridded baseline yields and estimates of yield impacts at N, P, and K supply shocks of 25\%, 50\%, or 75\%. To estimate the resulting impacts herein, we linearly interpolate between these steps. Beyond 75\% we extend the linear relationship between the 50\% and 75\% shock levels. These stepwise interpolations account for non-linear crop yield--nutrient relationships \citep{vangrinsven2022}. As discussed extensively by the authors, the yield impact estimates cannot fully capture all impacts due to observational constraints and are therefore primarily an approximation of short-term effects, most robust where fertilizer use is high \citep{ahvo2023}. To estimate changes in national production of each crop, we first derive national level impact coefficients by calculating the baseline production and reduced production for each crop by multiplying crop yields by harvested areas from the Spatial Production Allocation Model (SPAM) 2010 v2r0 \citep{ifpri2020, yu2020}. Subsequently, we apply this impact coefficient to the national production of the respective crop in a given year. For the bloc-level pipeline, these coefficients were precomputed as bloc-year-crop-nutrient dictionaries that reflect the weighted composition of each bloc in each year.

Using static crop areas for 2010 neglects the possibility that cultivation zones within countries may have shifted toward more or less suitable regions, thereby affecting crops’ nutrient responses. While we expect this effect to be marginal given the relatively short observation period and the aggregate scale of analysis, it should be borne in mind when interpreting the results.

\subsection{Crop trade, consumption, and calorie supply}

Bilateral trade data, domestic consumption, and opening stocks for the 11 staple crops (see above) were sourced from FAOSTAT for the years 1992-2023 \citep{faostat2024}. CEPII-BACI for natural gas required no reconciliation, while FAOSTAT crop and fertilizer bilateral flows were harmonized using a CIF/FOB parity correction ($=1.09$) with averaging of bilateral reports where both reporter perspectives were available. The 1.09 factor is the standard WTO/UNCTAD convention for CIF-to-FOB adjustment \citep{gaulier2010}.

Jointly, the selected crops provide around 70\% of the global human vegetal calorie intake \citep{faostat2024} and are representative of staple foods across major world regions including key tropical and temperate crops \citep{ahvo2023}. To estimate impacts on food security, we convert dry matter weights to calories based on conversion coefficients provided in \citet{faostat2024}. We acknowledge that this is a limited nutritional indicator neglecting crops' contributions to protein, vitamin, minerals, and dietary fiber supply but deem it sufficient for food security under disruption scenarios that pose a vital threat.

To evaluate the role of stocks and domestic crop production and their potential buffering capacity (see \autoref{fig:stockduration} and associated text), we divide the total calories in stocks for the considered crops by total calories embedded in their annual consumption.


\subsection{Panel data analysis}

Countries' ability to cope with agronomic input and food supply disruptions depends in part on their prevailing food security and overall socio-economic situation. That is, countries with commonly high calorie supply may be less affected by the same absolute loss than countries with low supply, and countries with greater economic means can more readily mitigate disruptions as long as international markets provide alternative supplies. Whether such mitigation measures can be implemented also depends on the nature of the disruption, as armed conflicts and destruction of physical trade infrastructures are far more challenging to compensate than trade conflicts. Therefore, we do not account for these factors endogenously in the modelling cascade but contextualize our findings \textit{ex post} using panel data. These comprise Gross Domestic Product (GDP) per capita based on purchasing power parity \citep{worldbank2025}, geographic regions \citep{worldbank2025}, World Bank country income groups \citep{fantom2016}, and the FAO prevalence of moderate to severe food insecurity \citep{faostat2024}.

\subsection{Estimation of centrality measures}

We assembled a directed, weighted multiplex of bilateral trade covering 208 countries, 15 commodity layers and 32 years (1992--2023). The layers comprise eleven staple crops, three macro-nutrient fertilizers, and natural gas. Trade flows are recorded in physical units -- tonnes for crops and fertilizers and billion cubic metres for natural gas -- and each link records the annual export volume from one country to another in a given commodity. Very small reported flows were dropped before analysis.

We characterized each country's role in each commodity layer and year using three complementary network centrality measures commonly used in academic literature -- \emph{Betweenness centrality} (framed in the text as \textit{routing role}), \emph{PageRank}  (framed as \textit{demand side}), and \emph{Eigenvector centrality} (framed as \textit{supply-core}) \citep{fagiolo2010,zhang2023,fagiolo2023}. Because raw trade volumes span many orders of magnitude, link weights were log-transformed and rescaled within each layer-year before centralities were computed, so that hubs in sparse layers (such as natural gas) and dense layers (such as wheat) can be compared on a common footing. All three scores were then rescaled to the unit interval. Implementation details are given in \autoref{app:formalization}.

To track how unequally trading power is distributed within each commodity layer over time, we summarized the country-level distribution of each centrality measure with the Gini coefficient, a standard one-number index of inequality ranging from zero (all active countries equally central) to one (a single country concentrating the entire centrality mass). Because the Gini's theoretical ceiling depends on the number of active countries -- which varies from roughly 17--40 active gas-trading countries to roughly 180 active crop-trading countries -- we applied a small-sample bias correction so that values from sparse and dense layers can be compared on a common 0--1 scale. Confidence bands were estimated by resampling countries with replacement (500 bootstrap replicates per layer-year-metric cell) and reporting the central 95\% of the resulting distribution. Per-layer Ginis were aggregated into commodity-type time series (crops, individual fertilizers, natural gas) using a weighting scheme that favors layers with more active countries. As a directly interpretable complement, we additionally report the share of total centrality held by the five leading countries in each layer-year. Mathematical definitions are in \autoref{app:formalization}.


\backmatter





\phantomsection
\addcontentsline{toc}{section}{Declarations}
\section*{Declarations}

\bmhead{Competing interests}
The authors declare no competing interests.

\bmhead{Data availability}
All data used in the study are publicly available from the cited repositories. Natural gas trade data are sourced from CEPII-BACI \url{https://www.cepii.fr/DATA\_DOWNLOAD/baci/doc/baci\_webpage.html}, while production and consumption data are provided by the Energy Institute \url{https://www.energyinst.org/statistical-review}. Data on the trade and production in crops and fertilizers, as well as food stocks, are available from FAOSTAT \url{https://www.fao.org/faostat/en/}. Harvested area data from the Spatial Production Allocation Model (SPAM) can be found at url{https://essd.copernicus.org/preprints/essd-2020-11/}. The dataset regarding crop yield impacts from changes in fertilizer input is available at \url{https://doi.org/10.5281/zenodo.8381197}. Income classification is available from the World Bank at \url{https://datatopics.worldbank.org/world-development-indicators/the-world-by-income-and-region.html}. Maps were created using Natural Earth 10m admin-0 product, available at \url{https://www.naturalearthdata.com/downloads/10m-cultural-vectors/10m-admin-0-countries/}. The code used to analyze these data and produce the results is available at \url{https://gitlab.iiasa.ac.at/anfos-public/food-disruptions-code}.

\phantomsection
\addcontentsline{toc}{section}{References}
\bibliography{sn-bibliography}

\newpage
\begin{appendices}
\setcounter{page}{1}
{\Large
\textbf{Supplementary text}}

\section{Country-level role profiles in market concentration by centrality measure}
\label{app:centrality}

Each country carries a distinct functional position in the directed weighted trade graph: routing (represented by \textit{betweenness centrality}) ranks transit and routing hubs, i.e. countries on shortest trade paths between others; demand-side (represented by \textit{PageRank centrality}) in our exporter to importer convention, ranks demand-side influence, i.e. countries receiving goods from highly connected partners; supply core (represented by \textit{eigenvector centrality}) ranks supply-core membership, i.e., in how far countries are embedded in densely-coupled exchange clusters. The same country can occupy very different positions across the three indicators (\ref{tab:country-roles}).

The United States is the most universal routing super-node: it ranks in the top 5 by betweenness in 13 of 15 trade layers and first for wheat, maize, sorghum, soybeans, nitrogen, and natural gas. Its other roles are narrower -- top 5 by PageRank in 5 layers and by eigenvector in 4 (first for sorghum, soybeans, and groundnuts) -- so the country dominates trade routing far more than demand-side influence or supply-core membership. The Netherlands is the single most multiply-central country and the prototypical demand and re-export hub: it ranks top 5 by PageRank in 9 layers -- first for wheat, maize, rice, sorghum, and groundnuts -- and by betweenness in 10, while also entering the supply core of all three fertilizers (eigenvector top 5 for N, P, and K). Germany is as a balanced multi-role node and a genuine supply-core anchor, ranking first by eigenvector for wheat, barley, and sugarbeet and in the top 5 across all three measures for several temperate crops. France is the perennial European runner-up, ranking top 5 by all three measures simultaneously for maize and sorghum and second by eigenvector for wheat, maize, barley, sorghum, and potatoes, yet rarely first. Russia is the clearest supply-core specialist, ranking first by eigenvector for all three fertilizers (N, P, and K) while appearing in only three betweenness layers and a single PageRank layer (potatoes). China couples an upstream routing role (betweenness rank 1 for phosphorus, potassium, and groundnuts) with the position of principal demand sink for natural gas (PageRank rank 1), but it is no longer a fertilizer supply-core leader (eigenvector rank 5 for both nitrogen and phosphorus). The United Kingdom is a demand-side specialist, ranking top 5 by PageRank in six layers -- chiefly cereals -- but by eigenvector in only two.

Temporal trajectories sharpen these roles. The United States has been the gas network's dominant routing hub throughout (betweenness at or near its maximum from 2008 to 2022, easing to 0.6 in 2023), but its demand-side role inverted: PageRank held at the ceiling through 2016 and then fell steadily (0.8 in 2018, 0.2 in 2022, 0.4 in 2023) as the country pivoted from a leading gas importer to a net exporter following the LNG build-out from 2016. Russia is the nitrogen supply-core anchor, its eigenvector centrality pinned at 1 every year since 1992; and its routing and demand-side roles peaked in 2021 (betweenness 0.7, PageRank 0.4) and then receded through 2022--2023. Ukraine shows the opposite pattern in wheat: its supply-core embedding rose to the structural maximum (eigenvector 1) in 2022 and 2023 -- by this measure the first two war years left it more central to the wheat export core. Morocco's grip on phosphorus has loosened, its eigenvector centrality easing from 1 in the early 1990s to 0.9 in 2023, with routing and demand-side roles declining in parallel. Finally, Turkey's nitrogen demand-side centrality climbed to its record high in 2023 (PageRank 0.5, up from 0.3 in 2016), consistent with the country absorbing displaced fertilizer flows as a major importer.


\section{The role structure of the energy--fertilizer--food trade network}
\label{app:netcomms}
\subsection*{Methodological note}
For each commodity -- the four single-layer fertilizer and natural-gas networks and an eleven-layer crop aggregate produced by per-layer log-mean normalization -- we constructed a directed supra-graph whose vertices are the participating $(\text{country},\text{year})$ pairs -- those that trade the commodity in a given year -- since a non-trading country-year carries no information and would otherwise dominate the change statistics. Year-$y$ trade edges carry layer index $y$ and log-weight $\log(1 + w_{ij})$, and inter-year edges coupling each country between its \emph{consecutive active} years carry a single dedicated coupling layer with constant weight. We inferred the partition with \texttt{graph-tool}'s \texttt{LayeredBlockState} \citep{peixoto2014b,peixoto2015} in two stages: a topology-only nested SBM (degree-corrected, best-of-five \texttt{minimize\_nested\_blockmodel\_dl} restarts followed by MCMC equilibration) was fitted, then warm-started a flat \texttt{LayeredBlockState} carrying the real-normal $\log(1+w)$ edge covariate, which was equilibrated and sampled to obtain $200$ posterior draws of $b_i(y)$.

From the MAP partition we computed per-year-pair node-flip counts as the Hamming distance between adjacent year-slabs, reporting the posterior mean and standard deviation across the 200 samples; per-country swing rates as the posterior-mean fraction of year-pairs in which a country's block label changed; and per-block dwell-time distributions as the run-lengths of identical block ids along each country's MAP trajectory. We further characterized the partitions with the adjusted mutual information between two commodities' partitions on their shared active set (cross-commodity role alignment), and the co-classification of geopolitical-bloc members over the network-wide baseline (bloc alignment).

\subsection*{Findings}

A dynamic stochastic block model fitted to the year-coupled supra-graph of each commodity resolves well-determined structure in every layer and year. The partitions expose a sharp architectural split (\ref{fig:sbm_structure}a). Crop, nitrogen, phosphorus, and potassium trade assign countries to the \emph{same} structural roles -- their partitions are mutually aligned at \textit{adjusted mutual information} $\mathrm{AMI}=0.48$ ($95\%$ posterior CI $[0.47,0.48]$, 2016--2023) -- so a country's role in crop trade closely predicts its role in fertilizer trade. Natural gas is functionally not reduced only to food sector, which is demonstrated by its structural decoupling from this complex ($\mathrm{AMI}=0.09$, $[0.08,0.12]$): its role structure shares little with food and fertilizer trade. Energy and food are thus not structurally merged but joined through a single narrow channel -- the gas-dependent synthesis of nitrogen.

The two systems are organized on different principles (\ref{fig:sbm_structure}b). Within the agri-food complex, structure is regional, and this holds across crops and all three fertilizers alike: members of regional trade blocs co-classify far more than chance -- in crop trade GCC, ASEAN, EAEU, MERCOSUR, EU, with the fertilizer layers enriched comparably -- whereas cross-regional political groupings are weaker. Natural gas shows much weaker bloc alignment: it is organized by export--import role and by pipeline and LNG infrastructure, cutting across geopolitical and regional blocs.

The systems also differ in temporal stability. The agrifood role structure is well-resolved each year but non-persistent, re-wiring $83$--$100\%$ of members between 1992 and 2023 (mean block dwell $1.3$--$1.7$~years; \ref{fig:rupture_all_commodities}) and continuously re-forming, with no secular consolidation into fewer roles (the number of roles is stationary across the record). Natural gas is, in contrast, a stable, low-dimensional role structure (three roles, mean dwell $6.3$~years) anchored by a persistent Russia--Caspian exporter core (Russia, Turkmenistan, Kazakhstan, Uzbekistan, Azerbaijan), reorganizing only episodically -- its two largest realignments coinciding with the commissioning of US LNG export capacity (2016) and the compounded epidemic and Ukraine-war disruptions (2020--2022).

Taken together, the latent role structure independently corroborates the architecture documented in the main text, which identifies a single, regionally-organized, continuously re-forming agrifood trading complex, fed by a structurally separate and infrastructure-bound energy system through the narrow, gas-dependent nitrogen bottleneck -- the upstream chokepoint whose disruption propagates downstream into food supply. The headline statistics are robust over the full record (agrifood alignment $\mathrm{AMI}=0.45$ versus gas $0.16$, regional-bloc enrichment $3$--$8\times$); all aggregates carry $95\%$ posterior credible intervals from the $200$ samples.

\section{Formalization}
\label{app:formalization}

The shock-propagation model cascades disruptions sequentially through three coupled trade networks: natural gas, fertilizers (N, P, K), and crops.  Two parallel pipelines were executed at country and bloc level, both terminating in a compounded caloric supply loss $\mathcal{L}^{\mathrm{final}}_{e,t}$ (Eq.~\ref{eq:final-loss}) used in the main text.  Throughout this section, $e$ indexes the analytical entity (a country in the first pipeline, a bloc in the second), $c$ a crop, $n \in \{\mathrm{N}, \mathrm{P}, \mathrm{K}\}$ a fertilizer nutrient, and $t \in \{1992,\dots,2023\}$ a calendar year.

\subsection*{Crop autarky shock}
For each entity~$e$, crop~$c$, and year~$t$, the \emph{crop autarky shock} was defined as the fraction of caloric supply that domestic production cannot satisfy:
\begin{equation}\label{eq:crop-autarky}
  \sigma^{\mathrm{crop}}_{e,c,t}
  = \max\!\Bigl(0,\;
      1 - \frac{P^{\mathrm{crop}}_{e,c,t}}{S^{\mathrm{crop}}_{e,c,t}}
    \Bigr) \times 100,
\end{equation}
where $P^{\mathrm{crop}}_{e,c,t}$ denotes domestic crop production (kcal) and $S^{\mathrm{crop}}_{e,c,t}$ denotes total supply (kcal), defined as production plus net imports.  In the bloc-level pipeline, production and supply were first summed across all member countries of bloc~$b$ in year~$t$ before applying Eq.~\eqref{eq:crop-autarky}.

\subsection*{Fertilizer autarky shocks}
An analogous autarky shock was computed for each fertilizer nutrient $n \in \{\mathrm{N}, \mathrm{P}, \mathrm{K}\}$:
\begin{equation}\label{eq:fert-autarky}
  \sigma^{\mathrm{fert}}_{e,n,t}
  = \max\!\Bigl(0,\;
      1 - \frac{P^{\mathrm{fert}}_{e,n,t}}{S^{\mathrm{fert}}_{e,n,t}}
    \Bigr) \times 100,
\end{equation}
where production and supply are measured in tonnes.

\subsection*{Natural gas autarky shock}
For natural gas, the autarky shock was defined identically but with consumption $C$ as the demand baseline:
\begin{equation}\label{eq:ng-autarky}
  \sigma^{\mathrm{NG}}_{e,t}
  = \max\!\Bigl(0,\;
      1 - \frac{P^{\mathrm{NG}}_{e,t}}{C^{\mathrm{NG}}_{e,t}}
    \Bigr) \times 100.
\end{equation}

\subsection*{Natural gas--to--nitrogen propagation}
Because ammonia synthesis is the dominant industrial use of natural gas, a shortfall in gas supply is transmitted to nitrogen fertilizer availability.  We modelled this dependency as an additive coupling: the natural gas autarky shock augments the nitrogen fertilizer deficit directly, subject to a ceiling of 100\%:
\begin{equation}\label{eq:ng-propagation}
  \hat{\sigma}^{\mathrm{fert}}_{e,\mathrm{N},t}
  = \min\!\bigl(100,\;
      \sigma^{\mathrm{fert}}_{e,\mathrm{N},t}
      + \alpha \cdot \sigma^{\mathrm{NG}}_{e,t}
    \bigr),
\end{equation}
where $\alpha = 1$ encodes the assumption that the entirety of the natural gas shortfall translates into an equivalent reduction in nitrogen fertilizer availability.

\subsection*{Crop response to single-nutrient shortages}
\label{sec:single-nutrient}
Country- and crop-specific retention coefficients were obtained by interpolating the spatially-explicit crop response curves from \citet{ahvo2023d}.  For each nutrient $n$, the single-nutrient retention coefficient at integer shock level $s$ is
\begin{equation}\label{eq:single-nutrient-lookup}
  r^{n}_{e,c,t}
  = \mathcal{R}^{n}\!\bigl(e,\,c,\,
       \mathrm{round}(\sigma^{\mathrm{fert}}_{e,n,t})\bigr),
  \qquad n \in \{\mathrm{N}, \mathrm{P}, \mathrm{K}\},
\end{equation}
where $\mathcal{R}^{n}: (\text{country}, \text{crop}, \text{shock level}) \to [0,1]$ encodes the response surface for a reduction in nutrient~$n$ alone, with the other two nutrients held at baseline; $r=1$ indicates no production loss and $r=0$ complete loss.  The corresponding crop production loss was scaled to the supply baseline so that it may be added to the autarky shock without double counting:
%
\begin{equation}\label{eq:single-nutrient-loss}
  L^{n}_{e,c,t}
  \;=\; \max\!\Bigl(0,\;
        \Bigl(1 - \frac{P^{\mathrm{crop}}_{e,c,t}}{S^{\mathrm{crop}}_{e,c,t}}\,
        r^{n}_{e,c,t}\Bigr) \times 100 \Bigr)
        \;-\; \sigma^{\mathrm{crop}}_{e,c,t}.
\end{equation}
%
For net importers ($P^{\mathrm{crop}}_{e,c,t} \le S^{\mathrm{crop}}_{e,c,t}$) the floor is inactive and Eq.~\eqref{eq:single-nutrient-loss} reduces to the supply-referenced production loss $\frac{P^{\mathrm{crop}}_{e,c,t}}{S^{\mathrm{crop}}_{e,c,t}}\bigl(1 - r^{n}_{e,c,t}\bigr) \times 100$, which together with the autarky shock $\sigma^{\mathrm{crop}}_{e,c,t} = \max\!\bigl(0,\, 1 - P^{\mathrm{crop}}_{e,c,t}/S^{\mathrm{crop}}_{e,c,t}\bigr) \times 100$ shares the same denominator and combines additively.  The $\max(0,\cdot)$ floor handles net exporters ($P^{\mathrm{crop}}_{e,c,t} > S^{\mathrm{crop}}_{e,c,t}$): when post-shock production $P^{\mathrm{crop}}_{e,c,t}\, r^{n}_{e,c,t}$ still exceeds domestic supply, the nutrient shock erodes only the export surplus, so the caloric supply loss is zero rather than $\frac{P^{\mathrm{crop}}_{e,c,t}}{S^{\mathrm{crop}}_{e,c,t}}\bigl(1 - r^{n}_{e,c,t}\bigr) \times 100$, which would over-count by that surplus.  For nitrogen we computed both an un-augmented version, $L^{\mathrm{N,no\,NG}}_{e,c,t}$, using $\sigma^{\mathrm{fert}}_{e,\mathrm{N},t}$ in Eq.~\eqref{eq:single-nutrient-lookup}, and an augmented version, $L^{\mathrm{N}}_{e,c,t}$, using $\hat{\sigma}^{\mathrm{fert}}_{e,\mathrm{N},t}$ from Eq.~\eqref{eq:ng-propagation}.

\subsection*{Natural gas contribution to crop loss}
\label{sec:ng-contribution}
The marginal effect of the natural gas channel on crop production is the difference between the two single-nutrient nitrogen losses defined in Eq.~\eqref{eq:single-nutrient-loss}:
\begin{equation}\label{eq:ng-contribution}
  \Delta^{\mathrm{NG}}_{e,c,t}
  \;=\; L^{\mathrm{N}}_{e,c,t} \;-\; L^{\mathrm{N,no\,NG}}_{e,c,t}
  \;.
\end{equation}

\subsection*{Aggregation to entity--year totals}
Crop-level losses were aggregated to entity--year totals by summing absolute caloric losses and dividing by total caloric supply:
\begin{equation}\label{eq:aggregation}
  X_{e,t}
  = \frac{\displaystyle\sum_c X^{\mathrm{kcal}}_{e,c,t}}
         {\displaystyle\sum_c S^{\mathrm{crop}}_{e,c,t}}
    \times 100,
\end{equation}
where $X^{\mathrm{kcal}}_{e,c,t} = S^{\mathrm{crop}}_{e,c,t} \cdot X_{e,c,t} / 100$ for any crop-level loss component $X \in \{\sigma^{\mathrm{crop}},\, L^{\mathrm{N,no\,NG}},\, L^{\mathrm{P}},\, L^{\mathrm{K}},\, \Delta^{\mathrm{NG}}\}$.  This supply-weighted scheme preserves the additivity of the components at the aggregate level: because every term shares supply as denominator, the entity-level decomposition retains the same arithmetic structure as the crop-level one.
In the bloc-level pipeline, production and supply were first pooled across the member countries of bloc~$b$ in year~$t$ at the autarky-shock stage (Eqs.~\eqref{eq:crop-autarky}--\eqref{eq:ng-autarky}); the aggregation in Eq.~\eqref{eq:aggregation} then sums the resulting crop-level caloric losses over crops within each bloc--year, exactly as in the country-level pipeline.

\subsection*{Compounded final loss}
\label{sec:final-loss}
The compounded final caloric loss reported in the main text combines the four independently-evaluated channels by which trade dependencies translate into caloric loss -- foreign supply (autarky), and single-nutrient nitrogen, phosphorus, and potassium shortages -- with the marginal natural-gas amplification on top:
\begin{equation}\label{eq:final-loss}
  \mathcal{L}^{\mathrm{final}}_{e,t}
  = \sigma^{\mathrm{crop}}_{e,t}
  + L^{\mathrm{N,no\,NG}}_{e,t}
  + L^{\mathrm{P}}_{e,t}
  + L^{\mathrm{K}}_{e,t}
  + \Delta^{\mathrm{NG}}_{e,t},
\end{equation}
where each term is the supply-weighted aggregate (Eq.~\ref{eq:aggregation}) of its crop-level counterpart: the autarky shock (Eq.~\ref{eq:crop-autarky}); the supply-rescaled single-nutrient losses for nitrogen without NG, phosphorus, and potassium (Eq.~\ref{eq:single-nutrient-loss}); and the natural gas contribution (Eq.~\ref{eq:ng-contribution}).  Each channel is supply-referenced so the five components combine additively without capping or rescaling.
Substituting Eq.~\eqref{eq:ng-contribution}, the last two terms collapse to the augmented single-nitrogen loss $L^{\mathrm{N}}_{e,t}$, so Eq.~\eqref{eq:final-loss} is algebraically equivalent to $\sigma^{\mathrm{crop}}_{e,t} + L^{\mathrm{N}}_{e,t} + L^{\mathrm{P}}_{e,t} + L^{\mathrm{K}}_{e,t}$; the five-term form is retained to make the natural gas contribution explicit in subsequent analyses.

\subsection*{Implementation details}

Both country- and bloc-level pipelines were implemented in \texttt{Python~3}.  Data manipulation used pandas and \texttt{NumPy} \cite{harris2020}. Spatial coefficient data were stored in a GeoPackage file and read via \texttt{GeoPandas} \cite{jordahl2020}; bloc-level coefficients were serialized as a \texttt{Python} pickle dictionary.  Piecewise-linear interpolation of retention coefficients was performed over four stages ([0,\,25], [25,\,50], [50,\,75], [75,\,100]\%) using \texttt{NumPy}'s linear interpolation routine, yielding integer-resolution lookup tables (101 entries per country--crop--nutrient combination). The binding shock level was rounded to the nearest integer before lookup. All intermediate caloric quantities were computed in double-precision floating point.

\subsection*{Integration index}
To quantify the degree to which geopolitical blocs trade preferentially among their own members, we computed a size-corrected integration index for each of the 20 blocs across all 15 commodity layers. For each bloc $b$ with $n_b$ member countries, we first extracted the layer-normalized directed trade flows for 2016–2023 from the multiplex adjacency matrices and partitioned them into three components: intra-bloc flow $F_\text{intra}$ (both endpoints in $b$), import flow $F_\text{import}$ (destination in $b$, origin outside), and export flow $F_\text{export}$ (origin in $b$, destination outside). The raw intra-bloc share $s_b = F_\text{intra} / (F_\text{intra} + F_\text{import} + F_\text{export})$ is, however, biased by bloc size: a 54-member bloc such as the African Union occupies a larger fraction of all possible country pairs and would therefore exhibit a higher $s_b$ than a 5-member bloc with identical per-pair trade intensity. We therefore normalize by the expected intra-bloc share under a null model of non-preferential, size-proportional mixing, $\hat{s}_b = n_b / N$, where $N = 208$ is the total number of countries in the multiplex. The integration index is defined as $\mathcal{I}_b = s_b / \hat{s}_b$; values $\mathcal{I}_b > 1$ indicate that the bloc trades more internally than its size alone would predict, whereas $\mathcal{I}_b <1$ indicates relative openness to extra-bloc partners. This index is invariant to bloc size, enabling direct comparison across blocs ranging from 3 (e.g. NAFTA) to 190 (e.g. Non-nuclear weapon bearing) members.

\subsection*{Centrality measures}

For every (year, layer) pair with at least three active countries we constructed a directed graph from the adjacency $A^{(\alpha, y)}$ and computed three centrality measures. To make heavy-tailed trade volumes commensurable across layers we first applied the per-layer log-transform
\[
\widetilde{A}_{ij} \;=\; \frac{\log(1 + A_{ij})}{\overline{\log(1 + A)}\big|_{A > 0}},
\]
dividing each present edge by the mean of $\log(1 + A)$ over the layer's non-zero entries; this places non-zero entries on a layer-comparable, mean-one scale.

\textit{Betweenness centrality} $C^{B}_{i}$ was computed as the share of weighted shortest paths through node $i$, with edge distance set to $d_{ij} = 1 / (1 + \widetilde{A}_{ij})$ so that high trade volume corresponds to short distance.

\textit{PageRank} $C^{P}_{i}$ was computed with damping factor $d = 0.85$, using $\widetilde{A}$ as the weighted transition propensity. In the exporter-to-importer convention $i \to j$, $C^{P}$ ranks demand-side influence: countries absorbing flow from highly-connected partners receive higher scores.

\textit{Eigenvector centrality} $C^{E}_{i}$ was obtained by power iteration on the normalized weighted adjacency $\widetilde{A}$, $\mathbf{v}_{k+1} = \widetilde{A} \mathbf{v}_{k} / \|\widetilde{A} \mathbf{v}_{k}\|_{\infty}$, initialized at $\mathbf{v}_{0} = N^{-1}\mathbf{1}$ and terminated at $\|\mathbf{v}_{k+1} - \mathbf{v}_{k}\|_{\infty} < 10^{-10}$ or $k = 300$. Power iteration was preferred over absolute-tolerance eigensolvers, which do not converge reliably on small-magnitude eigenvectors of normalized adjacencies. Each component was finally rescaled by its (year, layer) maximum so that scores lie in $[0, 1]$. Betweenness and PageRank were computed using the OpenMP-parallelised C++ routines of \texttt{graph-tool} \citep{peixoto2014b}.

\subsection*{Concentration of centrality across countries}

For each (year, layer, metric) triple, we summarized the country-level distribution of centrality with the bias-corrected Gini coefficient. Letting $v_{(1)} \le v_{(2)} \le \cdots \le v_{(n)}$ be the sorted non-negative centrality scores of the $n$ active countries, the raw Gini
\[
G_{\mathrm{raw}} \;=\; \frac{2 \sum_{i=1}^{n} i \, v_{(i)} \;-\; (n + 1) \sum_{i=1}^{n} v_{(i)}}{n \sum_{i=1}^{n} v_{(i)}}
\]
has theoretical ceiling $(n - 1)/n$, biasing it downward on layers with few active countries (natural gas, $n_{\mathrm{active}} \approx 17$--$40$) relative to crops ($n_{\mathrm{active}} \approx 180$). We therefore applied the small-sample correction from \citet{deltas2003},
\[
G_{\mathrm{adj}} \;=\; \frac{n}{n - 1} \, G_{\mathrm{raw}},
\]
so that $G_{\mathrm{adj}} \in [0, 1]$ on every layer regardless of network size; $G_{\mathrm{adj}} = 0$ corresponds to equal centrality across active countries and $G_{\mathrm{adj}} \to 1$ to one country concentrating the entire centrality mass. The statistic is undefined when $n < 2$ or when all scores are zero, the latter case arising for the natural-gas betweenness in years without any transit hub.

Confidence intervals were obtained by country-resampling bootstrap with $B = 500$ replicates: in each (year, layer, metric) cell we drew $n$ countries with replacement from the active set, recomputed $G_{\mathrm{adj}}$, and reported the 2.5th and 97.5th percentiles of the bootstrap distribution as the $95\%$ confidence band. For type-level aggregates (crops, single fertilizers, natural gas), the Gini was first computed per constituent layer and then combined across layers within a type using weights $w_{\alpha} = n_{\alpha}(n_{\alpha} - 1)$, which favor layers with larger active sets and reflect the dyadic count of pairwise comparisons underlying the Gini; the weighted-mean confidence band was obtained by propagating the per-layer bootstrap samples through the same weights before taking percentiles. As an interpretable complement bounded on $[k/n, 1]$, we additionally report the top-$k$ share $T_{k} = \big(\sum_{i=1}^{k} v_{(n - i + 1)}\big) / \sum_{i} v_{(i)}$ with $k = 5$, computed and bootstrapped identically.

\section{Methodological discussion}

\subsection*{Limitations in shock estimation}
\label{app:limits}

Setting a 1:1 natural gas supply deficit propagation assumes that all domestic N fertilizer production depends entirely on natural gas as a feedstock. In practice, a fraction of global ammonia synthesis uses coal gasification or electrolysis, particularly in China. This assumption represents an upper bound on natural gas dependency and may overstate the contribution of natural gas to caloric losses for countries with diversified ammonia feedstocks. In addition, the natural gas autarky shock used consumption as the denominator from \cite{ei2023}, whereas crop and fertilizer autarky shocks used total supply from \citet{faostat2024} and \citet{cepii2025}. This reflects differences in the underlying data conventions: natural gas statistics report consumption directly, while trade-flow-derived crop and fertilizer datasets define supply as production plus net imports. The two formulations are conceptually equivalent -- both measure the fraction of demand unmet by domestic production -- but the distinction should be noted when comparing shock magnitudes across tiers.

Crop yield impacts caused by losses in fertilizer supply were adopted from a public repository and an associated recently published study \citep{ahvo2023,ahvo2023d}. These yield impacts were estimated in the original study by training a Random Forest model on yield observations and fertilizer inputs around the year 2000. As the authors acknowledge, such models perform best near the center of the distribution and more poorly at the fringes, while extrapolation is not feasible. This renders the model constrained by and subject to the training data distributions. The authors nonetheless found a good fit relative to yield observations and responses occurring under nutrient input loss. Overall, their approach reflects short-term rather than long-term impacts under a new equilibrium, which would typically be more severe once soil nutrient stocks are depleted. \citet{vangrinsven2022}, for example, have comprehensively reviewed N responses of wheat under different baseline conditions. As the applied model is purely data-driven -- based on yield observations and spatially heterogeneous combined nutrient inputs without consideration of the underlying mechanisms in plant nutrition -- the input shocks of individual nutrients aggregate to the approximate total fertilizer shock provided by the authors of the original study. When no crop-response coefficient was available for a country--crop--nutrient combination, the fertilizer-induced loss was set to zero (i.e., full retention assumed). This is a conservative choice that may understate losses for countries with incomplete coefficient coverage in the \cite{ahvo2023} dataset.

The limitations discussed here may in part be overcome by using large-scale process-based crop models that simulate crop development and yield as a function of daily weather, agro-environmental conditions (e.g., soil, topography), and management practices. However, such models commonly have only N cycles implemented and, in some cases, P, whereas K still presents a challenge to simulate across varying crops and environments. Future research may therefore work toward harnessing the advantages of both data-driven and process-based modelling approaches to maximize the robustness of impact estimates.

Besides nutrient inputs, crop yields depend on a range of other inputs and would accordingly be affected by disruptions in those as well. \citet{ahvo2023} also included energy supply (translating into machinery use) and pesticides. These typically entail larger uncertainties regarding the spatial distributions of machinery use and how it translates into labor effectiveness, as well as the spatial use of pesticides and their effectiveness, both of which require spatial disaggregation of national-scale data \citep{vittis2021}. In the case of pesticides, more complex supply chains are furthermore involved for raw ingredients and intermediates that can hardly be reflected across the wide variety of compounds and products presently in use. We also do not account for disruptions in seed supply, for which we assume production typically takes place in nurseries in proximity to the region of use.

Country-specific crop-response coefficients were derived from a single cross-sectional dataset and do not vary over the 32-year study period. Bloc-level coefficients vary over time because they reflect changing bloc membership, but the underlying per-country parameters remain static; therefore, structural changes in agricultural technology or cropping patterns within countries are not captured due to limited data.

\subsection*{Fertilizer and crop reserves}

Due to data limitations and uncertainties, we do not account for reserves in our modelling chain, although they can buffer the impacts of supply shocks for some time. However, how this would play out in terms of production and export dynamics is a question of the conditions for the release of stocks that are typically maintained by (semi-)public institutions under government control. Since 2022, stockpiling of fertilizers has accelerated in countries that provide such information. Projections suggest that global fertilizer consumption will further increase after 2025, with much of this increase driven by strategic stockpiling rather than immediate agricultural needs \citep{international2025}. Earlier studies on food trade network disruptions have included food or crop commodity stocks and evaluated how they may buffer shock impacts under defined assumptions regarding their use and implications. Specifically, \cite{marchand2016} used a similar approach to the one employed herein for a cereal staple food trade network only, finding that, at the time of the study, reserves provided substantial buffering capacity in some countries but that there was an overall declining trend in reserves. Here, we estimate the duration of food stocks ex post for transparency, as their direct inclusion in disruption modelling would require further assumptions about their use. We acknowledge, however, that this would render our results less pessimistic for countries experiencing substantial shocks, at least in the short run, as we find that stocks in most regions suffice for only very limited periods (see \ref{fig:stockduration} and the associated text in the main body).

\subsection*{The role of trade infrastructures}
Our modelling of trade disruptions is based on transactions in goods among countries and various types of blocs but does not account for the physical dimension of their transport. For all commodities traded herein, maritime shipping is a key mode of international trade, for which maritime ports and shipping routes are essential infrastructure. The latter frequently require passage through straits and canals that have in the past posed chokepoints for international trade and can have tremendous impacts on international supply chains even in the case of local and regional conflicts. While their relevance has been quantified from an economic perspective in earlier research \citep{verschuur2025}, more granular data on commodities along transport routes and associated time series will be required to incorporate such scenarios into our approach.

\newpage
\section{Supplementary figures}
\begin{supplement}

\begin{figure}[H]
\centering
    \includegraphics[width=1\linewidth]{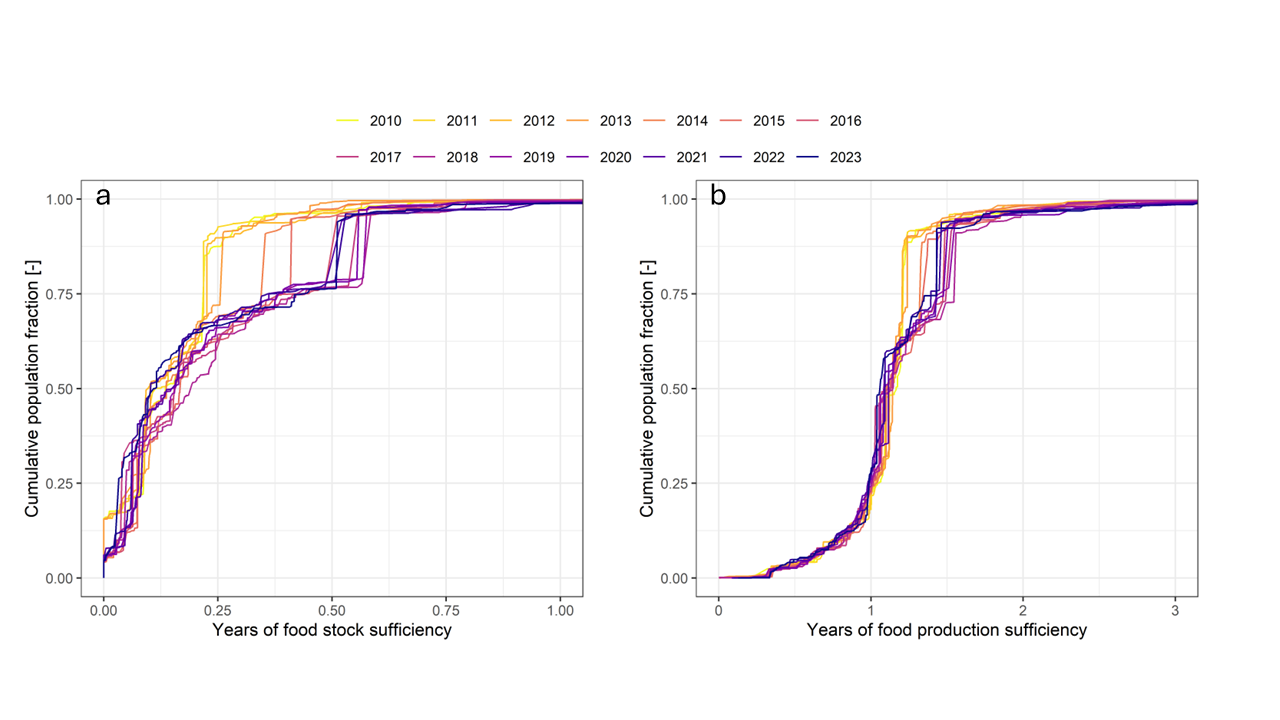}
    \caption{Same as \autoref{fig:stockduration} in the main body but showing annual population fractions.}
    \label{fig:stockdurationrel}
\end{figure}

\begin{figure}[H]
    \centering
    \includegraphics[width=0.9\textwidth]{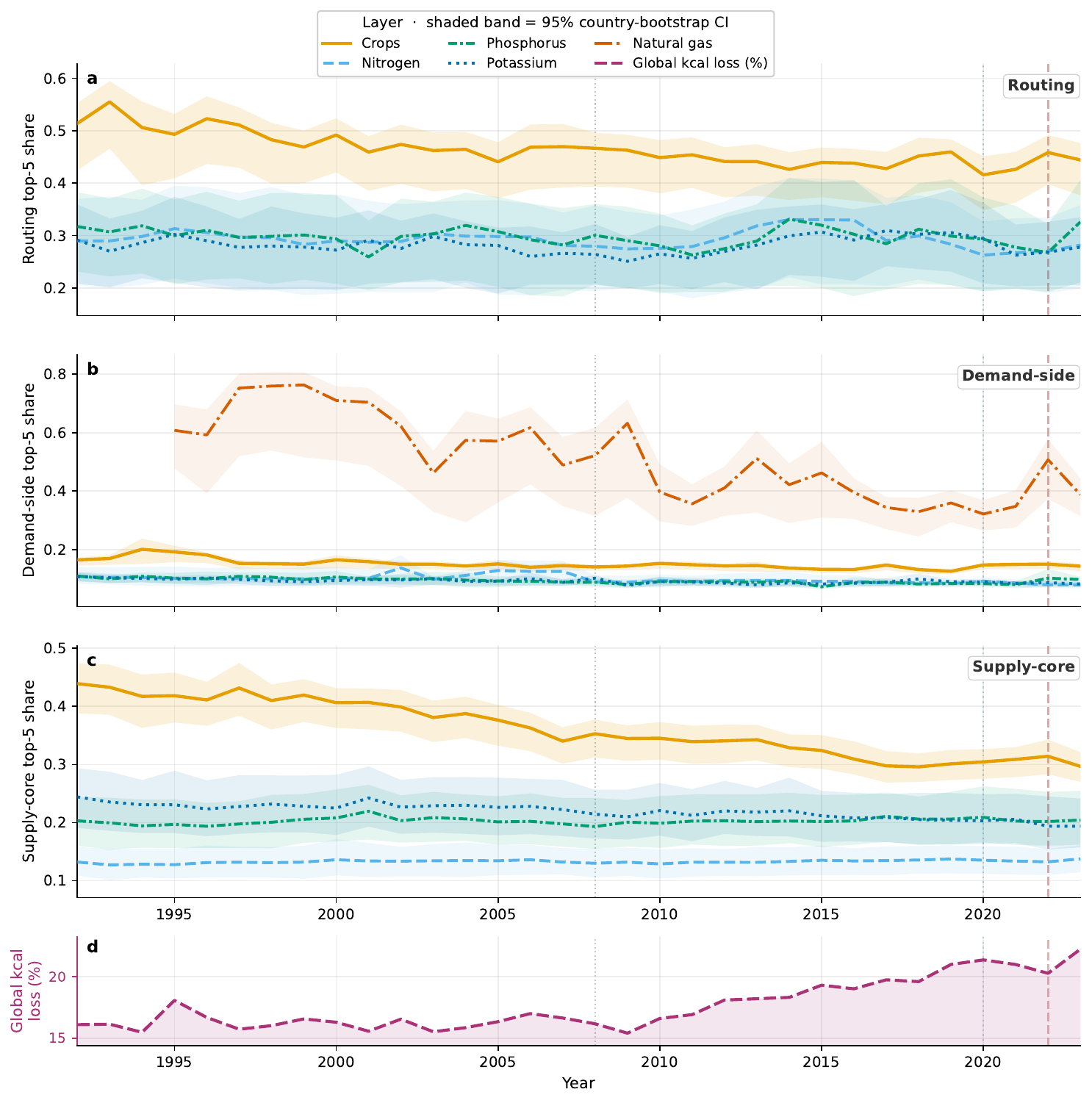}
    \hfill
    \caption{\textbf{Top-5 concentration of trade-network centrality, 1992–2023.} Top-K share (K = 5), defined as the fraction of total centrality held by the five leading countries in a given layer-year. Top-K share complements the Gini coefficient (\ref{fig:centrality}) as a size-robust concentration measure: unlike the Gini, its lower bound k/n depends only weakly on the number of active countries, making cross-layer comparisons between sparse (natural gas, $n \approx 17$--$40$) and dense (crops, $n \approx 180$) networks more interpretable.}
    \label{fig:topk}
\end{figure}

\begin{figure}[H]
\centering
\includegraphics[width=0.98\linewidth]{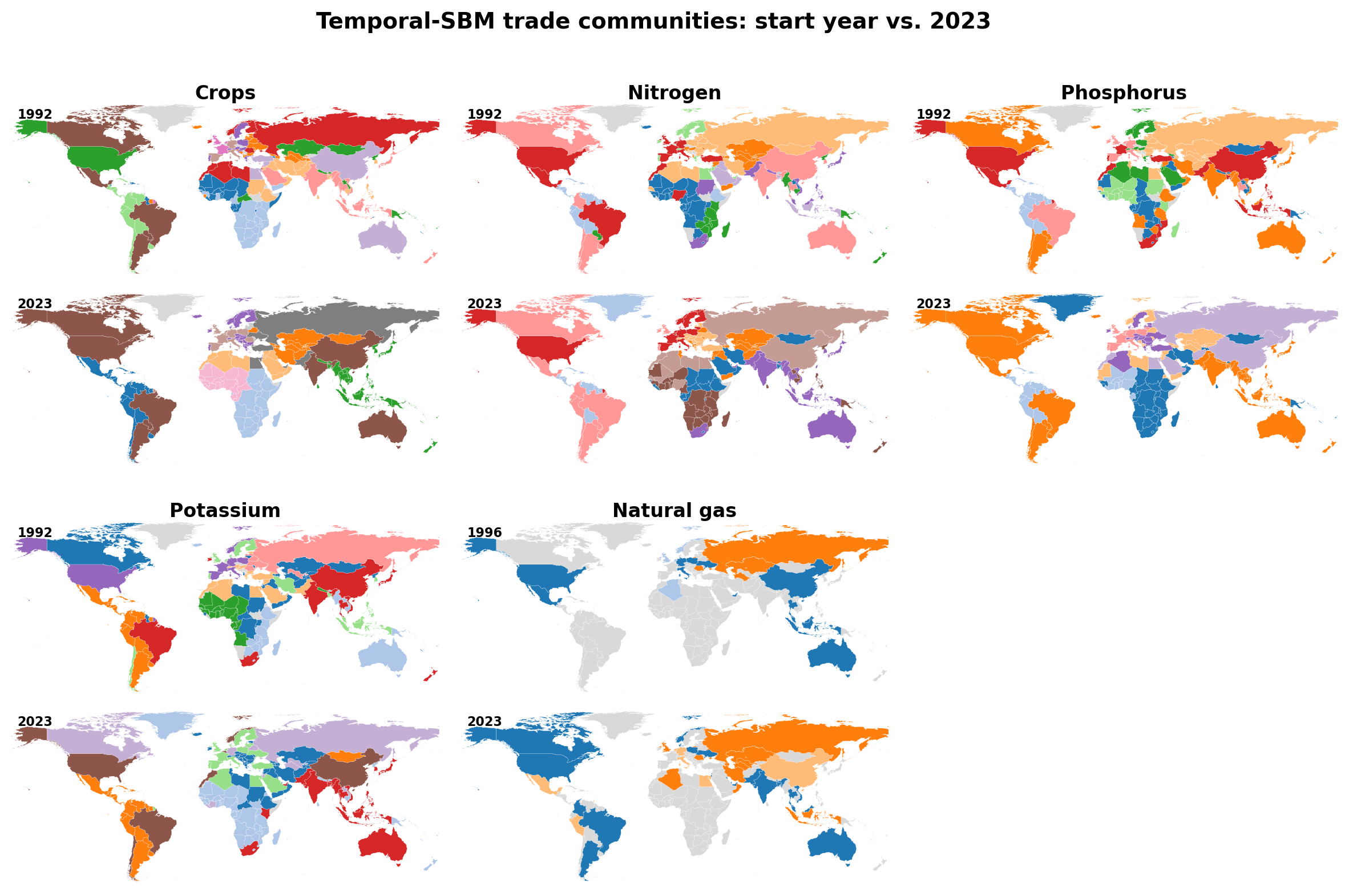}
\caption{\textbf{Temporal-SBM community structure of the five commodity-trade networks, 1992 vs.\ 2023.} World maps of country-level block assignments from the dynamic stochastic block model for the crop aggregate, the three fertilizer layers (nitrogen, phosphorus, potassium), and natural gas (shown 1996 vs.\ 2023, as gas trade is recorded in BACI only from that point). Each pair of sub-panels contrasts the partition at the start of the observation window with the partition at its end. Colors encode block identity within a commodity and are not comparable across commodities; within a commodity the labels are tied across years by the inter-year coupling edges, but because membership reshuffles almost completely in the food and fertilizer layers the cross-year color mapping is only indicative, and comparison should be made on the basis of \emph{which countries co-classify} rather than on color per se. The crop and fertilizer networks are well-resolved in each year but temporally non-persistent, recoloring extensively between 1992 and 2023 (\ref{fig:rupture_all_commodities}) -- the structure of globally integrated markets that continuously re-form. Natural gas represents a stable, low-dimensional role structure (three roles) anchored by at least one exporter core, whose membership largely preserved across the record.}
\label{fig:netcomms}
\end{figure}

\begin{figure}[H]
\centering
\includegraphics[width=0.95\linewidth]{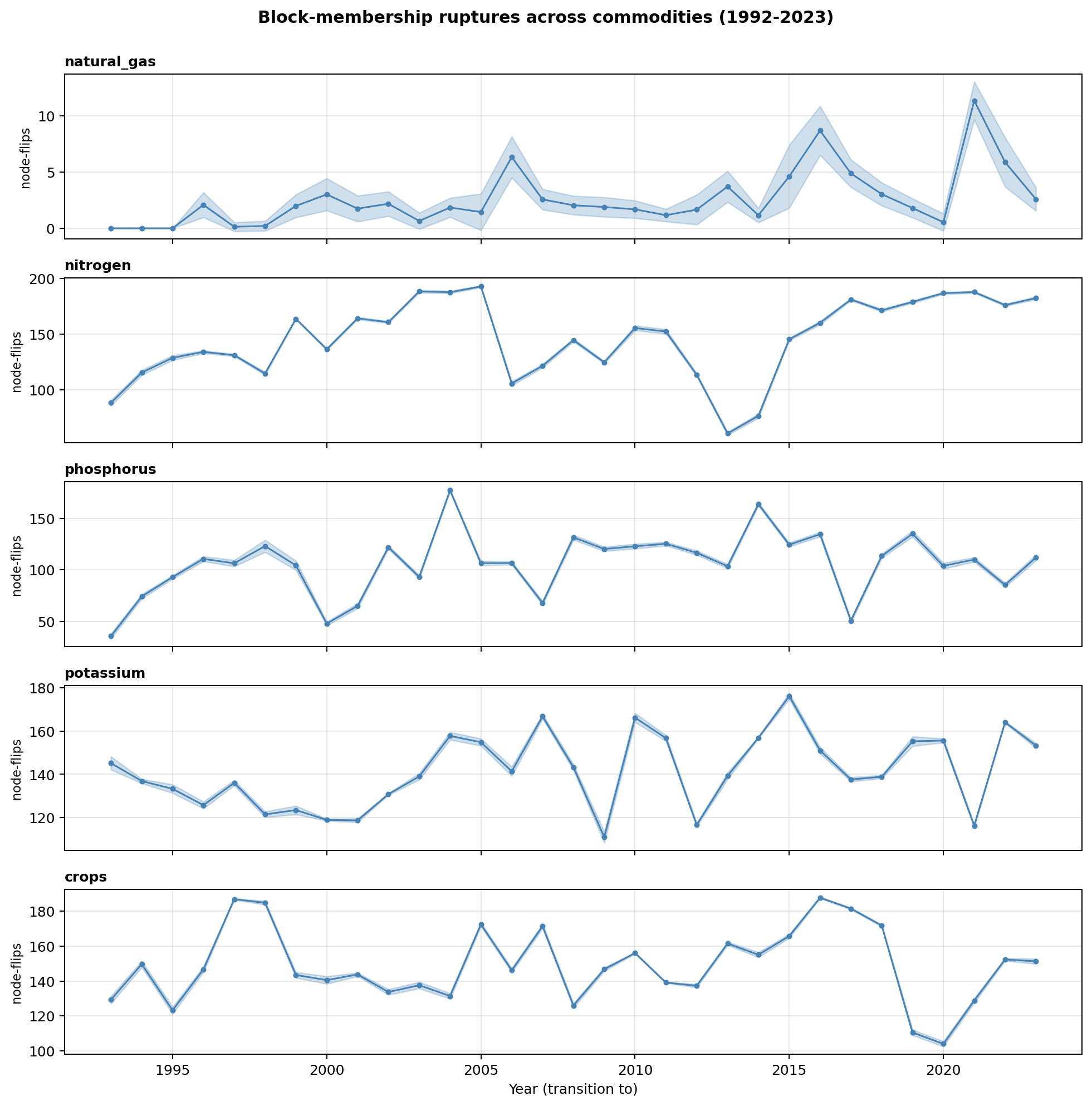}
\caption{\textbf{Block-membership ruptures across the five commodity networks, 1992--2023.} Posterior mean number of countries whose temporal-SBM block label changes between consecutive years (steelblue line and markers; shaded band, $\pm1$ posterior s.d.\ across $N=200$ MCMC samples) for natural gas, nitrogen, phosphorus, potassium, and the crop aggregate. The crop, nitrogen, phosphorus, and potassium networks show high background turnover, with no single year dominating: prominent peaks include nitrogen in 2002--2003 (188 flips) and 2022 ($176\pm1$), potassium in 2018--2019 (155 flips), and phosphorus in 2003--2004 (177 flips) and 2013--2014 (164 flips). Natural gas is structurally different: only $\sim\!40$ countries trade it in any year, and it sustains three stable communities, so it reshuffles rarely; its only pronounced reorderings falling in 2016 and 2020--2022.}
\label{fig:rupture_all_commodities}
\end{figure}

\begin{figure}[H]
\centering
\includegraphics[width=0.98\linewidth]{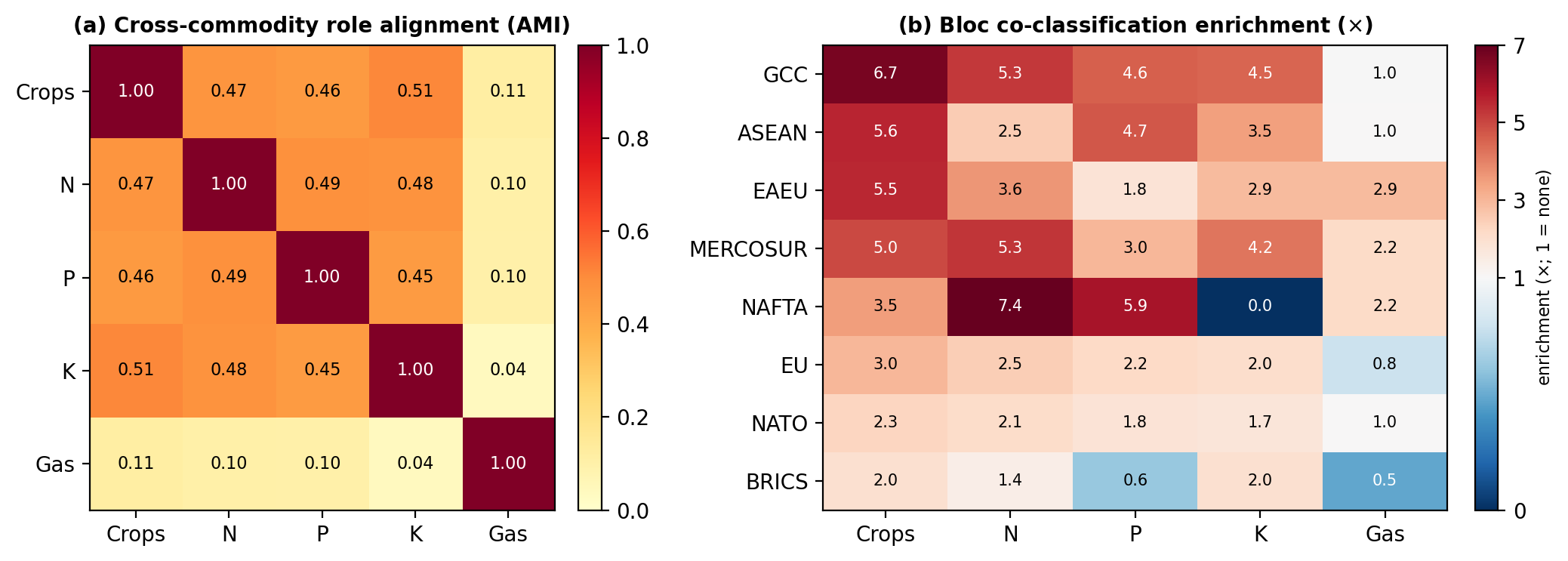}
\caption{\textbf{Latent role structure of the trade networks.} \textbf{(a)} Cross-commodity alignment of the temporal-SBM partitions (adjusted mutual information, 2016--2023 mean, on countries active in both layers): crop and fertilizer trade place countries into the same structural roles ($\mathrm{AMI}\approx0.45$--$0.51$), whereas natural gas is decoupled ($\mathrm{AMI}\approx0.09$). \textbf{(b)} Co-classification enrichment of geopolitical-bloc members over the network-wide baseline (2016--2023 mean; $1\times$ = no enrichment, red = enriched): the four agri-food layers (crops and the three fertilizers) are strongly enriched within the regional trade blocs ($\sim\!2$--$7\times$), whereas natural gas is not ($\approx\!1\times$) -- it is organized by export--import role rather than geography.}
\label{fig:sbm_structure}
\end{figure}

\newpage
\section{Supplementary tables}

\begin{table}[htbp]
  \centering
  \footnotesize
  \setlength{\tabcolsep}{4pt}
  \renewcommand{\arraystretch}{1.25}
  \caption{\textbf{Country roles in the food-energy-fertilizer trade by centrality measure (2016--2023 mean).} The three centrality measures pick out three different functional roles in the directed weighted trade graph: \emph{betweenness} ranks transit or routing hubs; \emph{PageRank}, in our exporter-importer convention, ranks demand-side influence; \emph{eigenvector} ranks supply-core membership. Each cell lists every layer in which the country reaches the layer's top 5 by the corresponding measure, with the rank shown as \textit{layer}(rank). ``---'' indicates the country is not in the top 5 by that measure. Tier I countries are top-5 in $\geq 5$ layers across multiple measures; Tier II countries concentrate in a single role; Tier III countries lead a single layer or a narrow specialty.}
  \label{tab:country-roles}
  \tiny
  \begin{tabular}{l p{1.5cm} p{2.7cm} p{2.7cm} p{2.7cm}}
    \toprule
    ISO & Country & Routing role: betweenness top-5 & Demand-side role: PageRank top-5 & Supply-core role: eigenvector top-5 \\
    \midrule
    \multicolumn{5}{l}{\textit{Tier I -- Multi-role super-nodes}} \\
    USA & United States  & N(1), gas(1), mai(1), soy(1), srg(1), wht(1), P(2), gnt(2), pot(2), rce(2), K(3), brl(3), sct(3) & sct(1), N(3), soy(3), brl(4), gas(4) & gnt(1), soy(1), srg(1), gas(2) \\
    NLD & Netherlands    & pot(1), sbt(2), gnt(3), srg(3), K(4), brl(4), gas(4), sct(4), P(5), soy(5) & gnt(1), mai(1), rce(1), srg(1), wht(1), brl(2), sbt(2), soy(2), pot(5) & pot(1), N(2), P(3), sbt(3), K(5), rce(5) \\
    DEU & Germany        & brl(1), sbt(1), K(2), wht(3), gnt(4), pot(4), srg(5) & sbt(1), soy(1), brl(3), gnt(3), rce(3), srg(3), wht(4) & brl(1), sbt(1), wht(1), K(2), N(3), pot(3) \\
    FRA & France         & brl(2), srg(2), wht(2), N(3), P(3), mai(3), pot(3), sbt(5) & K(3), mai(3), gnt(4), srg(4), N(5), soy(5) & brl(2), mai(2), pot(2), srg(2), wht(2), rce(4) \\
    CHN & China          & K(1), P(1), gnt(1), N(2), soy(3), rce(4), css(5), sct(5) & css(1), gas(1), K(4) & gnt(2), N(5), P(5) \\
    GBR & United Kingdom & gas(2), sbt(3), pot(5) & rce(2), srg(2), wht(3), mai(4), gas(5), sct(5) & brl(5), pot(5) \\
    \midrule
    \multicolumn{5}{l}{\textit{Tier II -- Single-role specialists}} \\
    RUS & Russia         & gas(3), N(4), wht(4) & pot(2) & K(1), N(1), P(1), gas(5) \\
    ITA & Italy          & rce(1), gas(5) & gnt(2), gas(3), rce(4), soy(4), srg(5) & rce(1), srg(5) \\
    BEL & Belgium        & -- & -- & P(2), sbt(2), K(3), N(4), pot(4) \\
    UKR & Ukraine        & -- & gas(2) & mai(3), soy(4), srg(4) \\
    BRA & Brazil         & -- & N(1), P(1) & soy(2) \\
    CAN & Canada         & soy(4), wht(5) & sct(3) & gas(1), soy(3) \\
    IND & India          & soy(2), gnt(5), rce(5) & -- & gnt(3) \\
    ARE & UAE            & rce(3), mai(4) & sct(4), brl(5), mai(5), wht(5) & -- \\
    \midrule
    \multicolumn{5}{l}{\textit{Tier III -- Layer specialists}} \\
    TTO & Trinidad and Tobago & -- & -- & gas(3) \\
    KAZ & Kazakhstan     & -- & pot(3) & gas(4) \\
    BLR & Belarus        & -- & -- & K(4) \\
    MAR & Morocco        & P(4) & -- & sct(1), P(4) \\
    ARG & Argentina      & -- & -- & mai(5) \\
    ROU & Romania        & -- & -- & mai(1), wht(4) \\
    \bottomrule
  \end{tabular}
  \begin{minipage}{0.97\linewidth}
    \footnotesize
    \raggedright
    Layer abbreviations: wht=wheat, mai=maize, rce=rice, brl=barley, srg=sorghum, soy=soyabeans, gnt=groundnuts, pot=potatoes, sbt=sugarbeet, sct=sugarcane, css=cassava, N=nitrogen, P=phosphorus, K=potassium, gas=natural gas. Within each cell, layers are listed in order of rank (rank 1 first) and ties broken alphabetically.
  \end{minipage}
\end{table}


\end{supplement}

\end{appendices}

\end{document}